# Laboratory and numerical study of intense envelope solitons of water waves: generation, reflection from a wall and collisions

A. Slunyaev[1], M. Klein[2], G.F. Clauss[2]

[1] *Institute of Applied Physics, 46 Ulyanova Street, Nizhny Novgorod, Box 120, 603950, Russia, Slunyaev@hydro.appl.sci-nnov.ru, and Nizhny Novgorod State Technical University n.a. R.E. Alekseev. 24 Minina Street, Nizhny Novgorod, 603950, Russia*
[2] *Ocean Engineering Division, Technical University of Berlin, Germany, MarcoKlein@email.de*

The investigation of dynamics of intense solitary wave groups of collinear surface waves is performed by means of numerical simulations of the Euler equations and laboratory experiments. The processes of solitary wave generation, reflection from a wall and collisions are considered. Steep solitary wave groups with characteristic steepness up to $kA_{cr} \approx 0.3$ (where $k$ is the dominant wavenumber, and $A_{cr}$ is the crest amplitude) are concerned. They approximately restore the structure after all the considered interactions. In the course of the interaction with the wall and collisions the maximum amplitude of the wave crests is shown to enhance up to 2.5 times. A standing-wave-like structure occurs in the vicinity of the wall, with certain locations of nodes and antinodes regardless the particular phase of the reflecting wave group. A strong asymmetry of the maximal wave groups due to an anomalous set-up is shown in situations of collisions of solitons with different frequencies of the carrier. In some situations of head-on collisions the amplitude of the highest wave is larger than in over-head collisions of the same solitons. The discovered effects in interactions of intense wave groups are important in the context of mechanisms and manifestations of oceanic rogue waves.

## I. INTRODUCTION

The group structure of sea waves is an important characteristic, which determines, for example, sailing conditions and wave loads on structures. The process of generation of waves by air is complex; the typical wind spectrum is not narrow, and the co-existence of different scales in the spectrum becomes apparent through the wave modulations. Due to the surface wave dispersion groups spread and bunch all the time in stochastic manner. Due to the four-wave nonlinear interactions, deep-water regular surface waves are prone to form groups. This effect is known as the modulational (or Benjamin – Feir) instability (Benjamin & Feir, 1967; Zakharov & Ostrovsky, 2009; Alber, 1978); though its significance in the real sea is still under discussion. The oceanic rogue waves are nowadays commonly related with the modulational instability, since it conduces to larger probability of large waves (Onorato et al., 2001, 2009; Kharif et al., 2009). The Benjamin – Feir instability is most efficient when the frequency spectrum is narrow. In particular, a crucial increase of the large wave probability when the spectrum is narrow has been observed many times in numerical simulations of unidirectional wave propagation within different frameworks, as well as in laboratory flume experiments (see e.g., Onorato et al, 2009; Shemer et al, 2009, 2010a,b). The modulational instability is evoked by the basic physical properties of deep-water gravity waves, the weak dispersion and nonlinearity; it is captured by the nonlinear Schrödinger (NLS) equation, which is the simplest framework for nonlinear modulated surface waves (Zakharov, 1968). The NLS equation for unidirectional waves is integrable and possesses envelope soliton solutions. These groups exist due to the balance between nonlinearity and dispersion; they do not disperse with time and do not disintegrate when interact with other waves, thus, are crucially different objects compared to the linear wave groups.

Though the modulational (Benjamin – Feir) instability originates from water waves, in the context of relatively steep surface waves envelope solitons have been believed for long to be a mathematical toy with no clear practical perspective. Steep soliton-shape groups were found to be structurally unstable in laboratory measurements by Yuen & Lake (1975): in contrast to groups with very small steepness $kA \approx 0.01$ (where $k$ is the wavenumber, and $A$ is the amplitude) groups with $kA \approx 0.2$ split into parts. Some more recent numerical simulations seemed to agree with that results (see review in Slunyaev et al., 2013a). In Ablowitz & Segur (1979) stable solitary wave groups with larger steepness $kA \approx 0.13$ were reported; their shapes agreed well to the NLS prediction. In publications by Dyachenko & Zakharov (2008), Slunyaev (2009) short solitary groups of very steep waves (close to the maximum Stokes wave steepness) were shown to propagate stably in numerical simulations of the potential Euler equations. In Slunyaev et al. (2013a) intense solitary groups with steepness $kA \approx 0.3$ were successfully reproduced in laboratory experiments. Though shapes and velocities of these stable groups differ from the analytic solution of the NLS equation, the persistence of the group shape is certainly in common between them. Moreover, the analytic NLS solutions could be used for generation of intense solitons by a wavemaker with an acceptable inaccuracy. Collisions of planer solitary wave groups were simulated within the Euler equations in Zakharov et al. (2006), Slunyaev (2009), and also by means of reduced Zakharov equations in Dyachenko et al. (2016). The general conclusion from these works is that the solitary groups interact to a great extent elastically.

Despite the fact that envelope solitons of water waves suffer from (at least) transverse instability (Zakharov & Rubenchik, 1974, Deconinck et al., 2006), they may represent important transitional dynamics of intense wave groups. Since in the weakly nonlinear weakly modulated limit the solitons are described by completely integrable NLS theory, they are advantageous for understanding of the complicated nonlinear wave behavior. When envelope solitons propagate over background waves of significant amplitude, they correspond to breather solutions of the NLS equation, which are prototypes of rogue waves (see, e.g., in Kharif et al., 2009). Due to particular combinations between the phases of solitary groups and the background, they may sum up and cause extremely large waves. Breathers have been already used in sea-keeping tests to model most dangerous (and most probable) sea wave groups (Onorato et al., 2013; Klein et al., 2016). The problem of interaction between waves and structures is of extreme practical interest, it has been extensively studied theoretically, numerically and experimentally, though the main interest was focused on the situations of regular, standing or solitary waves, which propagate over relatively shallow water (e.g., Chen et al. (2015) and references therein). The reflection of dynamically unstable groups was studied in recent works by Carbone et al. (2013), Viotti et al. (2014), Akrish et al. (2016), and the interest is growing.

The dynamics of soliton ensembles in different realms is in the focus of many recent studies. Intense solitary groups were revealed in time series of the sea surface displacement, which contained rogue waves, in Slunyaev et al., (2005, 2006). The development of optical fiber transmission lines demands deep understanding of the dynamics and statistical properties of envelope soliton ensembles. The evidence of presence of shallow water long-wave solitons in the in-situ data was presented in Costa et al. (2014). The appropriate combinations of solitary waves may lead to strong wave enhancement, and the understanding of this dynamics may provide further insights. In particular, pairwise soliton collisions were studied in Pelinovsky et al. (2013) for better understanding of the statistical properties of the soliton gaz. The construction of multisoliton wave fields which cause the biggest wave formation is discussed in the recent papers by Sun (2016) and Slunyaev & Pelinovsky (2016).

It is essential that the integrable NLS equation is valid for unidirectional waves with narrow spectrum, thus it cannot describe collisions of solitary groups with significantly different carrier wave vectors, including head-on and over-head collisions, when the interest is confined to collinear waves. The interacting wave systems with different carriers could be described by coupled NLS equations (Roskes, 1976; Ablowitz & Horikis, 2015, see also references in Onorato et al., 2006), though these systems are generally not integrable.

In the present work we further continue the investigation by Slunyaev et al. (2013a), where intense envelope solitons were reproduced in the laboratory flume. The maximum solitary groups were short (a couple of oscillations in a momentary wave profile); the waves were very steep, so that slightly larger initial conditions resulted in breaking of wave crests. The main objective of this work is to examine the process of collision of intense envelope solitons with a vertical wall, and to study interactions between intense envelope solitons, when the groups travel in the same or in opposite directions. The wall may model a side of a large boat, or a sea wall in a deep water area. As the number of laboratory runs is limited and the data which is acquired from the experiments is much incomplete, numerical simulations of the potential Euler equations are performed with the use of similar approaches as in Slunyaev (2009), Slunyaev et al. (2013a). The comparison between previous numerical simulations of intense groups and laboratory measurements (Slunyaev et al., 2013a,b) showed good agreement, thus the present simulations are considered to be trustworthy.

In Sec. II a brief general description of the laboratory facility and of the numerical method is given, and the way of preparing the wavemaker signal which produces a solitary group is discussed in Sec. III. Results of the laboratory experiments on generation and interactions of intense solitary groups are presented in Sec. IV; Sec. V contains the description of results of similar numerical simulations of the Euler equations. In conclusions we sum up the results of laboratory and numerical simulations and their impact on related issues.

## II. DESCRIPTION OF THE LABORATORY AND NUMERICAL APPROACHES

The laboratory tests were conducted in September, 2015, in the seakeeping basin of the Technical University of Berlin at conditions, similar to ones in the study by Slunyaev et al. (2013a). The basin is 110 m long, the width is 8 m and the water depth is 1 m. At one end, a fully computer controlled electrically driven wave generator is installed which was utilized in flap type mode. A nonremovable wave damping slope is installed on the opposite side of the tank. For the present experimental campaign a vertical wall was mounted before the absorber, made of oriented strand boards braced with wooden beams. The distance from the wavemaker to the wall was 95 m. The wall was surrounded by water from both sides; during the experimental tests on intense group reflection from the wall no movement of water from the other side of the wall was observed, what confirms the sufficient rigidity of the construction. In total, 30 runs were performed, only non-breaking cases were considered. The presence of the reflecting wall instead of absorbing beach made the measurements much longer, since it took time for waves to calm down.

Multiplication of the calculated wave sequence with the hydrodynamic as well as electric transfer function of the wave generator in frequency domain and subsequent inverse Fourier transformation result in the control signal for the wave generator (Biesel, 1951). The obtained control signal is afterwards checked against the wave generator limitations; see some more details on the procedure in Slunyaev et al. (2013a).

During the experimental campaign locations of the probes (9 gauges in total) were changed several times. Most of the probes were placed along the tank in its middle part, though in two setups (No 2, 3, see Fig. 1) a pair of gauges was placed in the cross section to check the uniformity of the transverse structure of wave groups.

The numerical simulations of water wave dynamics are performed by means of the High-Order Spectral Method (West et al., 1987) with 4-order Runge-Kutta integration in time, which solves the potential Euler equations for collinear waves. The scheme takes into account up to 7-wave nonlinear resonances (the nonlinearity parameter $M = 6$). Previously the code was verified versus laboratory measurements and versus simulations of other hydrodynamic models in Slunyaev (2009), Slunyaev et al. (2013a, b). For simulations of the nonlinear Schrödinger equation a standard pseudo-spectral method with split-step-Fourier integration in time is used.

## III. ENVELOPE SOLITON GENERATION IN THE LABORATORY BASIN

Envelope solitons are stable solutions of the nonlinear Schrödinger equation, which governs the evolution of a complex function of space and time, $A(x, t)$ (Zakharov, 1968; Hasimoto & Ono, 1972)

$$i\left(\frac{\partial A}{\partial t} + C_{gr}\frac{\partial A}{\partial x}\right) + \alpha|A|^2 A + \beta\frac{\partial^2 A}{\partial x^2} = 0. \tag{1}$$

The complex envelope $A(x, t)$ determines simultaneously the surface elevation $\eta(x, t)$ and the velocity potential at the water rest level, $\phi(x, t) = \varphi(x, z = 0, t)$ as

$$\eta = \mathrm{Re}(A\exp(i\omega_o t - ik_0 x)), \qquad \phi = -\frac{g}{\omega_0}\mathrm{Im}(A\exp(i\omega_o t - ik_0 x)), \tag{2}$$

where $k_0$ and $\omega_0$ are the wavenumber and cyclic frequency of the carrier, which are related according to the dispersion relation; $C_{gr}$ is the linear group velocity, $g$ is the gravity acceleration. In the case of an initial-value problem all the coefficients are functions of $k_0$ and water depth, $h$ (see Appendix, (A1)). In the limit of infinite water depth, $k_0 h \to \infty$, the dispersion relation and the coefficients tend to the following expressions,

$$\omega_0 = \sqrt{gk_0}, \qquad C_{gr} = \frac{\omega_0}{2k_0}, \qquad \beta = \frac{\omega_0}{8k_0^2}, \qquad \alpha = \frac{\omega_0 k_0^2}{2}. \tag{3}$$

In this limit the envelope soliton reads

$$A_{sol}(x,t) = \frac{A_0}{\cosh\left[\sqrt{2}k_0 a\left(x - C_{gr}t\right)\right]}\exp\left[i\frac{a^2}{4}\omega_0 t\right], \tag{4}$$

where dimensionless amplitude $a \equiv k_0 A_0$ characterizes the maximum wave steepness. The NLS theory and the envelope soliton solution may be reformulated for the boundary-value problem as described in Slunyaev et al. (2013a) or Chabchoub & Grimshaw (2016) (for the case of infinitively deep water). We do not reproduce here these details.

Within the integrable NLS equation (1) envelope solitons persist and interact elastically with all other waves including other envelope solitons. From the standpoint of laboratory experiments an envelope soliton is expected to be a structurally stable wave group, which does not exhibit significant variation of the envelope shape when it propagates. In our previous work (Slunyaev et al., 2013a) steep solitary wave groups were successfully reproduced in a wave flume. The steepest solitary groups were characterized by the dimensionless crest amplitude $A_{cr}\omega_m^2/g = 0.3$ (where $\omega_m^2/g$ estimates the wavenumber through the mean spectral cyclic frequency $\omega_m$, and $A_{cr}$ is the maximum crest amplitude in the group) and contained just a few wave cycles. The groups traveled for about 60 wavelengths or 15-30 group lengths along the tank without significant change of the group shape or noticeable radiation. Several methods for generating envelope solitons were examined in Slunyaev et al. (2013a) aiming at better reproduction of the solitons. The use of exact envelope soliton solution of the NLS equation was found to be efficient to generate the solitary groups of steep waves in the flume. At the same time, solution (4) does not describe the wave group asymmetry, neither the nonlinear correction to the group velocity. Though the instrumentally measured solitary groups do not coincide with solution (4), their physical features are much similar, thus in what follows they will be referred to as envelope solitons.

Based on the investigations in Slunyaev et al. (2013a), the following method is used for the reproduction of the solitons in the wave tank for this study. The method consists of a several steps. First, the exact solution (4) is used as an initial condition for the auxiliary numerical integration of the potential Euler equations for planar waves. In the course of evolution a stable solitary group emerges, while the radiated waves are damped by a co-moving mask (see Slunyaev et al. (2013a) for details). Thereby, the obtained time series of the surface elevation which correspond to the solitary groups of the Euler equations are converted to the wavemaker signal using the appropriate transfer functions. The strongly nonlinear solitary wave group is well specified in terms of the mean wavenumber, $k_m$ (in space domain), and mean frequency, $\omega_m$ (in time domain), and by the maximum crest and trough amplitudes, $A_{cr}$ and $A_{tr}$. In the present laboratory study carrier frequencies similar to the ones in our previous successful experiments are taken (Slunyaev et al., 2013a), see Table 1, 2. In some cases the appearance of new wave packets following the main wave group could be observed in the laboratory experiments due to the unwanted excitation of the second harmonic. The spectrum beyond frequency 10 rad/s was cut-off, what solved the problem.

The experimental facility has a depth $h = 1$ m and is generally not designed for simulations of deep-water waves, while the self-modulation effect which balances the group dispersion is strong in deep water. The influence of insufficiently deep water condition on the solitary groups is considered theoretically in Appendix. It is concluded, that even in relatively deep waters $kh \approx 5$ the solitons are noticeably different from those in the limit of infinite depth: for a given group width they possess effectively reduced amplitudes. The effect of wave group erroneous initial amplitude on the eventual amplitude of emerged envelope solitons is considered in Appendix analytically and by means of direct simulations of the NLS equation and the potential Euler equations. A moderate decrease/increase of the initial condition amplitude results in reduction/growth of the soliton amplitude with factor two. The results of these theoretical examinations were used for the laboratory experiments when adjusting the wavemaker signal. The dispersion relation and the wave group velocity related to the infinitively deep water condition (3) are employed in what follows, as they are weakly affected by the depth finiteness as discussed in Appendix.

## IV. LABORATORY EXPERIMENTS

At first, the experiments on single steep envelope soliton generation with parameters similar to the ones in Slunyaev et al. (2013) were repeated. It was found out that relatively short waves, $\omega_p \approx 7$ rad/s (Table 1, Exp. 1), approached the wall not perpendicularly, though no difference in the lengths of the flume sides was detected. Waves along the left-hand side of the basin (in the direction towards the mounted wall) traveled slightly faster. Consequently, the reflection from the wall was not exactly plane resulting in a strong steepening and even wave splashes in the far right corner of the tank. Moreover, the group seemed to start to disintegrate even before it approached the wall, and could completely lose the original structure at the wall, as shown in Fig. 2 for Exp. 1 (see records at probes 4-9, Fig. 1 represents locations of gauges in four series of experiments). Envelope solitons with longer carrier waves $\omega_p \approx 6$ rad/s (Table 1, Exp. 2, 3) turned out to be less affected by this fault. The waves were simultaneously recorded at two points of the tank cross section in two series of experiments: near the wall (Fig. 1, Setup 2) and about 15 m apart from the wall (Fig. 1, Setup 3). The corresponding records from Exp. 4, 5 shown in Fig. 2 agree satisfactory. Thus, frequency 6 rad/s was chosen for the study of the envelope soliton reflection from a vertical wall, though the water in this case is not really deep. Based on the experience gained in the previous study (Slunyaev et al. 2013a), the envelope solitons at frequency 6 rad/s are successfully produced in the present experimental campaign by adjusting the wave amplitude as discussed in Appendix.

Due to the transverse instability of planar envelope solitons (4), weak perturbations with lateral scale $k_y$ grow exponentially as

$$\sim \exp(ik_y y + t/T_{trans}), \quad k_y = \sqrt{2}ak_0\rho, \quad T_{trans} = \frac{4}{a^2\omega_0\Omega}, \tag{5}$$

where the dimensionless transverse wavenumber $\rho$ and the instability increment $\Omega$ are discussed in detail in Deconinck et al. (2006) within the NLS framework. According to Zakharov & Rubenchik (1974) in the limit of long transverse perturbations (small $\rho$) the scales of instability are related as

$$\Omega^2 \approx \frac{4}{3}\rho^2\left(1 - \frac{1}{3}\left(\frac{\pi^2}{3} - 1\right)\rho^2\right), \tag{6}$$

and thus most unstable perturbations exist and are most likely to grow first in laboratory simulations. An improved numerical analysis of the transverse instability of planar NLS envelope solitons by Deconinck et al. (2006) concluded that the most unstable lateral scale corresponds to $\rho \approx 0.8$ and $\Omega \approx 0.657$. When for conditions of the present laboratory experiments one takes $a = 0.3$, $\omega_0 = 6$–7 rad/s and the deep-water dispersion relation (3) is applied, then the most unstable scale corresponds to $L_y = 2\pi/k_y \approx 3.7$–5 m with characteristic time scale $T_{trans} \approx 10$–11 s. Thus, potentially the transverse instability could disturb the groups in our laboratory experiments, as it takes about 100 s for the groups to pass the basin. At the same time the visual observations seem to suggest that the dominant scale of the transverse variation of waves is equal to the double lateral size of the basin, 16 m, which is quite different from the estimation of the most unstable perturbation. Though, particular investigation of this issue was not performed. Therefore the physical reason of the observed unwanted effect is not identified with confidence.

When the groups propagate lonely, the group parameters may be evaluated on the basis of the few recorded time series (for Exp. 1-3 – probes 1-4 before reflection, for Exp. 4 – gauges 1-3 before reflection of the leading soliton, $t < 120$–140 s): the frequency of the spectrum peak, $\omega_p$, the mean frequency of the energy spectrum, $\omega_m$, the maximum crest amplitude, $A_{cr}$, and trough amplitude, $A_{tr}$. They are given in Table 1 versus the same quantities calculated from the data of auxiliary numerical simulations used for producing the wavemaker signal. The two wave groups in Exp. 4 correspond to the same conditions, and thus are analyzed jointly. The values from numerical and laboratory simulations agree rather well, though the amplitudes of laboratory groups seem to be smaller (since only a few measured time series of short wave groups are available, the maximum wave amplitudes are surely underestimated). In this range of water depths the dispersion relation is almost unaffected by depth, and the deep-water linear dispersion relation (3) may be used to estimate the mean wavenumer as $\omega_m^2/g$. The dimensionless depth parameters, $h\omega_m^2/g$, are also listed in Table 1. The corresponding frequency spectra for Exp. 3, 4 are shown in Fig. 3. The spectra at different locations seem to be very similar, what confirms the constancy of the wave group shape. On the other hand, the spectrum of steep solitary groups depends on the phase of the maximum wave as shown in Slunyaev et al. (2013), thus some variability of the spectrum is expected. By vertical dashed lines and dash-dotted lines frequencies $\omega_p$, $2\omega_p$, $3\omega_p$ and $\omega_m$, $2\omega_m$, $3\omega_m$ are shown correspondingly in Fig. 3 for Exp. 3, 4. The frequencies are calculated as the average among the values calculated at different probes. It may be noted that though the main spectral peaks are better described by $\omega_p$, the second and third harmonics, which are the phase-locked nonlinear modes, are better estimated with multiple mean frequencies $2\omega_m$ and $3\omega_m$.

Table 2 contains the most important parameters of experiments 5-7, when two solitary groups are different. All the parameters in Table 2 are calculated from the numerical simulations of the Euler equations. The mean wavenumber $k_m$ is obtained in the spatial domain on the basis of the wavenumber spectrum; due to nonlinearity it slightly differs from $\omega_m^2/g$. In Exp. 5 the groups from Exp. 4 and Exp. 3 are reproduced. In Exp. 6, 7 the groups are located very close to each other; the frequency spectra for Exp. 6 are shown in Fig. 3 at different locations for $t < 135$ s; the two-peak spectrum is clearly observed. Velocities of the solitary wave groups, which are obtained from the numerical simulations, are also given in Table 2. With use of these values, the wave group positions are tracked in experiments on soliton collisions in Fig. 2, see vertical lines with symbols on tops for Exp. 4-7 (records from gauge 1 are used to define the initial locations). Very good agreement for the distance over few tens of wave periods may be concluded, though some tendency to overestimate the velocity of the second soliton may be found in Exp. 4, 5. Thus, the numerically simulated solitary groups are rather similar to their laboratory counterparts.

**A. Reflection from vertical wall**

Almost perfect reflections of soliton envelopes from the vertical wall are observed in Exp. 2, 3. The reflected groups preserve their structure and continue to propagate with no clear tendency to deform, see records in Fig. 2. The steeper solitary group in Exp. 3 seems to deform faster than the smaller soliton in Exp. 2. In spite of large steepness of the waves, no overturning of individual waves was observed.

After the wavemaker launches the wavetrain, it returns to the vertical position and may serve as the second vertical wall in the flume. Therefore the solitary groups could experience more than one reflection travelling along the tank, though in the course of propagation the waves at one side of the tank delayed, what eventually resulted in distortion of the wave group structure and complete disintegration of the solitary group later on. This process is displayed in long records of Exp. 2, 3 (Fig. 2), see intervals $t > 150$ s. The groups split into parts after the second reflection from the motionless wavemaker.

Arrays of gauges were located in the vicinity of the wall in Exp. 1-5 with spacing about 0.5 m with the purpose to register the process of reflection in detail. In Exp. 1-3 and 5 probe No 9 was placed just 1 cm from the wall for registration of the contact point. Unfortunately a bias error of the data acquisition channel was revealed during the campaign, which occurred occasionally and could not be recognized during the calibration procedure. Therefore the amplitude of the records from gauge No 9 in Exp. 1-3 may have incorrect magnitude. One may notice from Fig. 2 that the records at gauge No 8 in Exp. 2, 3 (a half meter before the wall) exhibit much smaller surface displacements compared with the time series from gauges 6 and 7. The range of recorded displacements at different probes is shown with bars in Fig. 4. Some variability of the maximum displacement along the tank is observed (which may be partly due to the varying phase of the waves), which is definitively more pronounced in the vicinity of the wall. The doubtful data from probe 9 in Exp. 2, 3 are shown by interrupted strips in Fig. 4. The surface displacement near the wall may significantly increase. In Exp. 5 the maximum of the record from probe 8 (which is about 0.5 m from the wall) corresponds to a twofold enhancement, $\max(\eta(t))\omega_0^2/g \approx 0.63$. Records from some other gauges near the wall in Exp. 1-3, 5 may exhibit wave subsidence (e.g., gauge 8 in Exp. 2, 3).

## B. Head-on collision of solitons

The head-on collision between two envelope solitons was arranged with use of reflection from the wall. The wave maker launched a sequence of two wave groups with some delay between them. After the first group reflects, it experiences a head-on collision with the second group (Exp. 4, 5). Since the planar structure of solitary groups tends to break, the solitons were targeted to collide at shorter possible distance, not far from the wall at $x = 80$ m according to the linear theory.

In Exp. 4 the two groups are similar; they are steeper than in Exp. 2, and approximately same in steepness as the group in Exp. 3. The frequency spectrum for the groups from Exp. 4 is shown in Fig. 3, which is rather similar to the spectrum for Exp. 3.

In Exp. 5 the solitary groups are different in amplitude but have similar frequencies; individually they repeat the groups from Exp. 4 and Exp 3, see Table 1, 2. The phases of solitons in Exp. 4 and 5 are different.

In Exp. 4, 5 the leading soliton reflects from the wall and then collides with the second soliton. According to the visual observations during the experiments, the actual focusing point was not far from $x = 80$ m. The initial conditions in Exp. 4 and Exp. 5 are slightly different (in phases of the solitons and the amplitude of the second soliton). As a result, the records of surface displacements near the distance 80 m look qualitatively different in Exp. 4 and Exp. 5 (cf. in Fig. 2, 4). After the head-on collision the second soliton reflects from the wall, and follows the leading soliton towards the wavemaker. After the reflections the solitons start to disintegrate qualitatively similar to the observation in Exp. 3, when a single soliton propagated (see Fig. 2).

The gauges are installed at different locations in Exp. 4 and Exp. 5 (Setup 2 and 3 respectively, see Fig. 1). In Exp. 4 the transverse structure of the waves may be checked at about 1 m before the wall with the help of two gauges located at the center of the flume (gauge 6) and near the side-wall (gauge 7). The records from the gauges 6 and 7 in Exp. 4 look similar, see Fig. 2, thus the waves are almost uniform in the transverse direction. It is clear from Fig. 4 that the highest waves occur at the point of collision (gauge 4) and 1 m before the wall (gauge 8). At the gauge 9 which is located 50 cm from the wall the waves are, in opposite, smaller.

The significant wave enhancement which is observed in Exp. 5 1 m before the wall has been already mentioned. In this experiment gauges 4 and 5 measure waves close to the focusing point near the center and near the side-wall of the experimental basin. They show similar time dependencies when the first soliton passes (see Fig. 2 at $t \approx 105$ s). Approximate wave annihilation during the solitons' collision (Fig. 2, $t \approx 140$ s) is also recorded by the two probes. In fact, the 'annihilation' event is similar to a nodal point of a standing wave. When the second (steeper) soliton passes the point $x = 80$ m after it reflects from the wall (Fig. 2, $t \approx 170$ s), the difference between the time series from probes 4 and 5 becomes obvious, thus the group is no longer uniform in the transverse direction.

Similar to the process of reflection at the vertical wall, no wave breaking were observed in Exp. 4, 5 despite large wave steepness.

## C. Overhead collision of solitons

Contradictory conditions should be fulfilled for the purpose of simulation of the overhead collision. The carrier frequencies should have significant difference to provide relatively short time of collision caused by different velocities of the groups. Short wave groups were found to experience strong lateral deformation; they are worse reproduced by the wavemaker. Longer waves correspond to the shallower condition and thus envelope solitons for given steepness are longer. Longer groups of longer waves should be still consistent with the size of the flume.

Two experiments with compromise parameters are presented in this sub-section (Exp. 6, 7). The groups are characterized by two frequencies about 7 rad/s (the first group) and 6 rad/s (the second group); they have similar steepness $k_m A_{cr} \approx 0.20$, see Table 2, which is noticeably smaller than in the previous experiments. The reason for this was occasional breaking of the wave crests observed in the experiments with groups of the steepness ~0.3. Opposite wave phases are the only difference between the initial conditions for Exp. 6 and Exp. 7. The time series recorded by the gauges in Exp. 6 and Exp. 7 are plotted with different lines in Fig. 2. The combination of records in these experiments exhibits the conditional wave envelope in the time domain. This figure also demonstrates the repeatability of experiments. The frequency spectrum variation along the tank is shown in Fig. 3 for Exp. 6. Two main peaks are well seen in the figure. The second harmonics also may be recognized. During the collision between the solitons the spectral peaks stay separated all the time; some broadening of the spectrum may be noticed.

The interaction between solitons occurs before the waves approach the far wall of the tank. The leading group with smaller amplitude and shorter waves (gauge 1) propagates slower, and at gauge 9 appears behind the larger and faster group. The interaction process resembles the classic exchange interaction of solitons, the groups remain rather well isolated all the time. At the same time, larger waves occur during the interaction. It is interesting to note that the velocities of the solitary groups, obtained in the numerical simulations, are capable of accurate description of the soliton movement even when the solitons exchange, see Fig. 2 (Exp. 6&7). It is difficult to say from Fig. 2 whether the interaction has resulted in shifts of the soliton paths, as the classic NLS theory foresees.

When the groups pass one through the other and reflect from the wall, they pass the probes again ($t > 140$ s). The larger group of longer waves seems to be probably more stable at long time, than the smaller group of shorter waves. We can speculate that a similar effect was observed in numerical simulations of planar waves by Slunyaev (2009), where a collision of solitary groups with different carrier waves resulted in disintegration of the group with shorter waves within the Euler equation framework.

## V. NUMERICAL SIMULATIONS

Since the number of laboratory experiments is limited, and the information which is obtained by means of measurements of the surface displacement in a few points is much incomplete, supplementing numerical simulations of the Euler equations for collinear waves were performed by means of the HOSM. The comparison between intense solitary groups obtained in the numerical simulations and measured in the laboratory conditions showed very good agreement in Slunyaev et al. (2013a). The code uses periodic boundary conditions; with the purpose to model reflection by a vertical wall, mirror symmetric initial conditions are used. When waves propagate in opposite directions, they collide symmetrically at some point due to the periodic boundary condition. In the collision point the fluid cannot move horizontally due to the imposed symmetry, what is equivalent to a virtual vertical wall.

The parameters of experiments are chosen similar to the laboratory condition, the length of the numerical tank is $L = 90$ m, it is infinitively deep. A few reference cases are considered to obtain the general picture of the physical effects. The initial conditions for the numerical simulations are specified in a similar manner as the boundary condition in the laboratory experiments, i.e. stable wave groups obtained in the preparatory long-term numerical simulations of the Euler equations are used. In particular, solitary groups of two characteristic steepnesses, $\omega_m^2/g\, A_{cr} = 0.33$, $k_m A_{cr} = 0.30$ and $\omega_m^2/g\, A_{cr} = 0.15$, $k_m A_{cr} = 0.14$, are considered, which are respectively a very steep but still realistic condition, and a moderate steepness case when the group is not too long (the length of the envelope soliton (4) is inverse proportional to its amplitude). In the discussion below the soliton steepnesses will be approximated by values of 0.3 and 0.15 respectively. These solitary groups are shown in Fig. 5. For the purpose of scaling the typical wale length, $\lambda_m = 2\pi/k_m$, and the mean period $T_m = 2\pi/\omega_m$ will be used. In different simulations three carrier frequencies are considered, 6, 7 and 8.5 rad/s. The groups propagate about twice slower than the wave phase, thus the water displacement in a particular point depends on a combination of the location of the group and of the wave phase. For generality, initial conditions with at least six different wave phases are simulated in parallel, as shown by a series of curves in Fig. 5, one of the curves is given by a contrast line. Besides the surface elevations, the imagined enveloping curves are shown in Fig. 5, which are composed by local maxima/minima among the wave realizations with different phases. The upper and lower enveloping curves in Fig. 5b are noticeably different, what is due to the asymmetry of nonlinear waves.

### A. Reflection from vertical wall

Reflections of solitary groups with frequency 6 rad/s have been simulated. At moment $t = 0$ solitons are located at the left side of the numerical wave tank, $x = 0$. The distance to the other side of the tank corresponds to more than 50 wavelengths; it takes more than 100 dominant wave periods to reach the wall. The wave dynamics near the wall is shown in Figs. 6-8. In Fig. 6 the contours of the upper envelope are shown in the area of 10 wavelengths from the wall for 40 wave periods. Close to the wall a picture of interference is observed, which is more evident in the case of less steep waves, when the envelope is broader (Fig. 6a). In the area near the wall (with size of the characteristic width of the envelope soliton) the waves are locally standing waves; in particular, the first waist of the wave field is located at the distance about $\sim\lambda_m/4$ from the wall (slightly closer) regardless the particular wave phase. The limits of the wave displacement observed along the numerical tank during the simulations of waves with different phases are shown in Fig. 7 by the shading. Examples of the surface displacement which corresponds to the maximum rise of water at the wall are shown by curves. The most intensive displacements are localized within 1/4 of the wave length near the wall (see the vertical dashed line at $1/4\lambda_m$ from the wall in Fig. 7). The particular localizations of the wave maxima and minima in space and time in one simulation depend on the wave phase.

The time series of water displacement at the wall are given in Fig. 8. The maximum and minimum values of the displacements in the tank in all the six simulations for different wave phases as functions of time are shown by the shading. Examples of the time series at the wall are plotted as well by curves. Some noisy ripples after the reflection of the main group may be noticed in Fig. 8b, which indicate that the solitary group is strictly speaking not a true soliton. The broken horizontal lines in Fig. 8 show levels of the half of the maximum/minimum displacements which are attained at the wall during the reflection. It may be seen that when the waves are steep the maximum drop of the water level at the wall is about twice the deepest trough of the soliton in the both cases, while the highest rise of the water is noticeably larger than twice the crest amplitude of the soliton (Fig. 8b). The enhancements of the maximum wave crest in Figs. 8a,b are 2.17 and 2.46 respectively. Similar to the laboratory observations, the numerical simulations do not exhibit breaking phenomenon even in the case of a steeper soltion.

### B. Head-on collisions of envelope solitons

The reflection from a vertical wall is in fact a particular case of a head-on collision – of a solitary group with its virtual mirror reflection, when they are perfectly phase-matched. Then a standing wave appears locally in the vicinity of the wall as a result (Fig. 6). In the general case when phases of the interacting solitons are not related, the locations of wave nodes depend on the particular phase combination, and the standing-wave-like structure of the enveloping curves does not appear. The space-time diagram of the surface displacement for an example of a head-on collision is given in Fig. 9a (note that unlike Fig. 9, the envelope was plotted in Fig. 6). The coordinate is commoving with the velocity of the envelope soliton with longer carrier. The inclined variations of colour intensity along the soliton paths reflect the wave phases which propagate about twice faster than the groups do.

Head-on interactions between solitary wave groups with the same carrier frequency are presented in Fig. 10 by virtue of the maximum/minimum limits of the surface displacement in the entire simulation domain among all considered combinations of wave phases as functions of space/time (shading), and by the space/time series of the wave with maximum wave crest (thin lines). Collisions of solitons with different carrier frequencies are shown in a similar way in Fig. 11 (see the legend for head-on collisions). Interactions with a steep solitary group may result in occurrence of strongly localized enhanced waves, represented by 1-2 wave cycles in the space domain.

As could be expected, the maximum displacement due to the head-on collision is about the sum of the crest amplitudes of the interacting groups. The horizontal broken lines in Fig. 10, 11 show the half of the maximum positive and negative values of the data in a corresponding figure. The head-on collision of two similar solitons of moderate steepness corresponds to doubling of the wave amplitude (Fig. 10a). The head-on collision of two steep solitons is shown in Fig. 10c; this case is close to the situation represented by Fig. 7b, 8b; the wave crest enhancement is 2.34. In the case shown in Fig. 11b the amplitudes of the two solitons are about the same, thus the maximum wave amplitude is almost doubled during the collision. When the interacting groups have different amplitudes, the wave enhancement (i.e. the ratio of the maximum displacement during the collision over the maximum displacement of waves before the collision) is less than 2, see Fig. 10b.

## C. Over-head collisions of envelope solitons

An over-head collision formally may be considered as a head-on collision of waves in a system of references moving with an appropriate velocity, though then a background current exists which complicates the problem in nonlinear setting. The crucial difference is that if frequencies of the waves are of the same order of magnitude, then the process of interaction between envelope solitons which propagate in the same direction takes much longer time than in head-on collisions; thus nonlinear effects have favorable conditions to act. The carrier frequencies of solitons should differ significantly to provide relatively short time of interaction. Collisions of solitons with two values of steepness and three values of carrier frequency are examined in this subsection.

Two examples of over-head collisions are shown in Fig. 9b,c. The pairs of solitons in panels (b) and (c) have the same steepnesses, but the dominant frequencies are closer in Fig. 9c. Some shift of solitons' paths may be noticed (in particular, the path of the second (longer-wave) soliton shifts rightwards in Fig. 9c, the vertical dashed lines are added in the figures to assist in making this observation); this shift is more substantial when the solitons have close frequencies. Local areas of wave amplification/attenuation in time and space are very well seen in Fig. 9, though their positions strongly depend on the particular wave phases.

Different cases of over-head soliton interactions are shown in Fig. 11 by thin curves (surface displacement) and by shading with the dashed curve at the edges (limits of the displacements within the simulation domain). These results are compared in Fig. 11 with the simulations of head-on collisions of the same solitary groups (see the previous subsection). The shaded areas have significantly longer time scale in the over-head collisions. On the other hand, one may see that a head-on interaction of envelope solitons can yield somewhat larger wave amplification during the collision than the over-head interaction does (Fig. 11c,d). In the case displayed in Fig. 11d the crest amplification is 1.60 in the head-on collision versus 1.40 in the over-heading; in the case shown in Fig. 11c it is 1.36 versus 1.25 respectively. In collisions in Fig. 11c,d the steepest soliton has bigger amplitude compared to the other soliton. To some extent this observation contradicts the laboratory experiments reported in Sec. IV, as the solitons with steepness ~0.3 did not break in head-on collisions in the laboratory facility, but overturned in the tests of over-head interactions. A possible explanation of this inconsistency may be referred to the instability of the transverse structure of the steep solitary groups of short waves, which was observed in the laboratory flume. The long interaction with another packet might effectively increase the steepness of the solitary waves and further the lateral instability of the short-wave group, what could eventually cause local wave breaking.

It is interesting to note the difference between the maximum wave shapes when colliding solitons have the same (Fig. 10, head-on collisions) or different frequencies (Fig. 11, head-on and over-head collisions). The momentary profiles of colliding groups of different carriers gain anomalously high set-ups (cf. left columns of Fig. 10a and 11a; Fig. 10b and 11b,c; Fig. 10c and 11d). Of course, momentary shapes of the maximum waves in Fig. 10 (left column) do posses set-ups since they belong to the short groups. But the troughs in Fig. 10 touch (or almost touch) the lower edge of the shaded area, though it is not the case in Fig. 11, where the troughs are significantly higher than the lower edge of the shading. This peculiarity is practically absent in time series of head-on collisions, though the over-head interactions lead to a noticeable set-up in time series of the maximum waves as well (see corresponding curves in the right column of Fig. 11).

## VI. CONCLUSIONS

The dynamics of intense solitary packets of collinear waves is studied in the laboratory tank and by means of numerical simulations of the potential Euler equations: the reflection from a vertical wall, head-on and over-head collisions. In the laboratory tests, solitary groups are generated by a wavemaker at one side of the basin with a vertical wall at the opposite side; the wave evolution is recorded with the help of 9 probes, which are located appropriately. The initial-value problem in a periodic space domain is considered in the numerical simulations. The numerical simulations are repeated several times for different phases of the waves, what allows visualizing the wave envelope and obtaining a more general picture of the results incorporating various combinations of the wave phases. The laboratory measurement of the over-head collision of intense envelope solitons is also performed for two different wave phases.

The study is focused on steep wave conditions; the steepness of the simulated solitary groups in terms of the mean frequency and the crest amplitude is up to about $A_{cr}\omega_m^2/g \approx 0.3$, similar to the one in (Slunyaev et al, 2013a). No wave breaking is observed in the laboratory experiments on the reflection from the wall and the head-on collisions of solitary groups even in the steepest cases. Solitary groups with carrier frequency 7 rad/s exhibit some transverse nonuniformity, which develops before they reached the reflecting wall. As a result, locally larger wave steepness occurs and leads to a local breaking. The estimation of conditions for a transverse instability of NLS envelope solitons does not result in a certain conclusion. The longer waves with carrier frequency 6 rad/s are significantly more stable and hence are considered in the majority of the experiments. For this frequency the total length of experimental basin corresponds to about 55 wave lengths.

Other effects which can disturb formation of the wanted solitary groups are the effect of finite water depth and of the distorted amplitude of the boundary/initial condition. They are examined with the help of exact solutions of the approximate integrable nonlinear Schrödinger equation, and by means of direct numerical solution of the NLS and Euler equations. A moderate decrease/increase of the initial condition amplitude results in a reduction/growth of the soliton amplitude with factor two. For a given initial condition the effect of finite depth causes reduction of amplitude of the eventual solitary group. This effect is significant even when water is relatively deep: the dimensionless depth $kh \approx 5$ corresponds to a 20% drop down of the soliton amplitude with respect to the case of infinite depth. An envelope soliton does not emerge at all when $kh < 2$.

Interactions of solitary wave groups of collinear waves with a vertical wall and between each other are found to a large degree elastic in both, laboratory and numerical simulations. A shift of envelope solitons' locations caused by the interaction is observed in the numerical simulations of the Euler equations; it is more pronounced when the carrier frequencies are close. In the numerical simulations of collisions between pairs of solitons with steepness 0.3 the energy conservation error reached relatively large values, ~0.1%. The numerical instability of the High Order Spectral code, which took into account up to 7-wave interactions, was successfully regularized in most cases by a high-frequency filtering, though in physical experiments these situations might correspond to the onset of the wave breaking. In the situations of smoother waves the energy was preserved with a few orders of better accuracy.

An array of wave probes recorded the surface displacement in the area near the wall. Unfortunately, the reliable data of the surface displacement on the wall were collected in a single laboratory experiment (Exp. 5). The complementary simulations of the solitary wave group reflections were performed numerically; they foresee that the maximum water elevation on the wall (when waves do not break) may be up to about 2.5 times higher, and the deepest depression is about 2 times large than far from the wall. The effect of the wall on waves is equivalent to a mirror reflection, which leads to appearance of a standing wave structure of the wave envelope with maximum on the wall and the nearest waist at about one fourth of the dominant wave length from the wall. Depending on the wave phase, the particular location of the wave maxima/minima in time and space may vary. The local enhancement and subsidence of waves are registered in the laboratory experiments. Similar to the case of refraction, collisions of solitary groups may also exhibit local nodes and antinodes, so that the head-on colliding solitary groups may 'annihilate' at some locations, this effect is confirmed in both, numerical and laboratory simulations. However, the envelopes of colliding packets are smooth after averaging over possible wave phases, in contrast to the case of reflection on a wall.

Roughly speaking, the maximum possible cumulative wave amplitude is given by a superposition of the partial wavegroup amplitudes with a correction due to nonlinearity. When the maximum surface displacements before and during the interactions are compared, the biggest amplification is caused by collisions of waves with similar amplitudes. Some significant features of the solitary wave interactions are found in the numerical simulations: i) when the steepest soliton is the largest one, then the head-on collision results in a larger wave crest than the over-head collision; ii) when two interacting solitary groups have different carrier frequencies, the wave with maximum crest experiences an anomalous set-up. The set-up is more evident for the snapshots of the head-on collisions rather than for the time-series; it is even better seen for space- and time-series of the over-head collisions (Fig. 11).

The conclusions on the maximum wave, which may be observed when solitary wave groups collide, supplement the studies of the maximum wave due to the nonlinear modulational instability (Tanaka, 1990, Slunyaev & Shrira, 2013). The interest to specific shapes of extreme sea waves is of evident practical reason; it has motivated a series of recent studies (Adcock et al., 2011; Slunyaev & Shrira, 2013; Adcock, 2016). The discovered peculiarities of the interactions between nonlinear wave groups may give insight into the problem of interpretation of extreme wave records. The shapes of the maximum waves caused by soliton interactions are quite specific (Fig. 10, 11) and resemble some rogue wave shapes observed in our simulations of steep irregular waves (Sergeeva & Slunyaev, 2013). The set-up of the New Year Wave (not a set-down, as could be expected from the NLS theory) provokes much discussion about its origin, and, consequently, about the physical mechanisms of rogue wave generation; crossing sea conditions were suggested to explain this peculiarity (Adcock et al, 2011). In the present work this idea may obtain some further support.

The NLS equation and related analysis of the wave field with respect to the modulational instability in terms of the Benjamin – Feir Index (Onorato et al., 2001) are based on a narrow-banded spectrum and a weakly nonlinear approximations, and thus are frequently criticized by opponents, when applied to the real sea conditions. In this work we show though still somehow purified but realistic scenarios when NLS-like solitary groups can cause extreme wave patterns. These groups of very steep waves are able to propagate for a significant distance with some degree of firmness when collide with other waves and reflecting structures. The groups may consist of just a few oscillations, and thus are characterized by a wide spectrum. Collisions of solitary groups with different peak frequencies (multi-peak spectra) look consistent with expectations from the nonlinear Schrödinger equation framework.

## ACKNOWLEDGEMENTS

AS acknowledges support from Volkswagen Foundation (the laboratory experiment campaign) and RFBR grants 14-02-00983, 15-35-20563, 16-55-52019. The numerical simulations were performed within the RSF grant No 16-17-00041.

## APPENDIX: EFFECTS OF IMPERFECT CONDITIONS

An attempt to reproduce stable short solitary wave groups in a laboratory flume may fail due to a number of effects which change the initial/boundary condition or change the wave guiding properties of the basin. Two relevant problems which influence the amplitude of the generated solitary wave group are considered in the two sub-sections below: of incorrect wave amplitude and of insufficient water depth.

### 1. Effect of an erroneous amplitude

If a wave generator distorts wave/group shape, the actual dynamics of the group may differ significantly from expected. For example, Yuen & Lake (1975) could not observe the specific soliton dynamics in the situation of relatively steep waves ($a \approx 0.2$) as the generated at one side of the flume 'soliton' split into parts in the course of propagation. The characteristic scale of this effect depends on the wave steepness as $a^{-2}$, and is much larger (for wave of moderate steepness – $O(10^2)$ wave periods) than the scale of the emergence of multiple frequency packets. Two cases may be distinguished depending on whether dispersion overbalances the nonlinear coupling or the nonlinearity dominates. These situations may take place when the amplitude of the wavemaker signal is smaller or larger than necessary.

The effect of modified amplitude of the initial condition on the eventual emergence of a solitary wave group may be easily estimated within the NLS framework (1). The initial value problem for a group $A(x, t = 0) = \mu A_{sol}(x, t = 0)$, where $\mu > 0$ is a constant, was studied by Satsuma & Yajima (1974). According to their exact analytic solution the amplitude of the resulting envelope soliton is equal to $(2\mu - 1)A_0$. Thus, the deficit/surplus of the amplitude of the initial condition is transferred to the decrease/increase of the resulting soliton with a factor 2, see the dashed line in Fig. 12. We verify this theoretic estimation by means of direct numerical simulations of the potential Euler equations in deep water. To this end the Cauchy problem is simulated in a large spatial domain with periodic boundary conditions for the analytic NLS envelope soliton (4) with an amplitude factor $\mu$, $\mu A_{sol}(x, 0)$.

For example we consider a rather steep reference envelope soliton with a steepness $a = 0.25$. The numerical solution at large time is analyzed with the purpose to estimate the amplitude of the resulting solitary wave group. Due to the finiteness of the computational domain, the solitary group often propagates above radiated waves which spread along the domain. Besides, when the wave group is too short, its maximum amplitude depends on the wave phase and thus oscillates in time. To evade these problems, the maximum wave amplitude is estimated in statistical manner. Namely, the wave evolution for a few tens of wave periods it tracked, and the mean value of the maximum surface displacement and its standard deviation are calculated. The means and the intervals of standard deviation are shown by wide red boxes in Fig. 12 for the HOSM simulations. The same procedure is applied to the modulus of the complex envelope, $|A|$, and to the surface elevation, $\eta$, simulated by means of the NLS equation, see respectively the narrow blue boxes and circles with error bars in Fig. 12. The 3-order reconstruction formula is used instead of the leading order approximation for the surface displacement (2) (see e.g., in Slunyaev et al., 2013a). In case $\mu = 1$ the maximum of $|A|$ is exactly $A_0$, as must be. According to the solution of Satsuma & Yajima (1974) solitons do not emerge at all when $\mu < 0.5$; in this case waves basically spread along the available spatial domain. Assuming that the energy is distributed uniformly along the computational domain, a trivial estimation for the maximal achievable amplitude may be suggested, which is given by dotted line in Fig. 12. In general, one can conclude that a reasonable agreement between the simulations of the full and the approximate equations is observed in Fig. 12. The analytic estimation represented by the two lines captures the behavior of the simulated cases quite well.

The typical dynamics of solitary groups with modified amplitudes is shown in co-moving references in Fig. 13a ($\mu = 0.8$) and Fig. 13b ($\mu = 1.2$) for the NLS framework with the focus on the bottom parts of the groups. The wave group spreading is obvious in Fig. 13a, when the nonlinearity is not sufficient to bound the waves. In the opposite situation, when the nonlinearity overbalances dispersive effects (Fig. 13b), some wave radiation seems to appear as well, but in contrast to the previous case the emission is accompanied by observable beating between the solitary group and the radiated train due to a significant difference between their frequencies. Necks of the envelope are located at the both sides of the main group, they reveal the phase inversion of envelope $A$. This peculiarity helps to identify the situations of the amplitude excess or deficit and then to apply the following strategy. In the course of the laboratory experiments the recorded data were examined visually, and if necessary the wavemaker signal amplitude was adjusted to balance the nonlinearity and the dispersion.

## 2. Effect of finite water depth

In Slunyaev et al. (2013) steep and short envelope solitons were reproduced at conditions when the dimensionless water depth belonged to the interval $3.5 < kh < 5.8$. The available frequency domain for generated waves is determined by many physical and technical issues, therefore waves in the present study correspond to the similar values of depths. The coefficients of the NLS equation (1) depend on the depth according to formulas (e.g. Slunyaev, 2005)

$$\omega_0 = \sqrt{gk_0\sigma}, \quad \sigma = \tanh k_0 h, \quad C_{gr} = \frac{\omega_0}{2k_0}\left(1 + k_0 h \frac{1-\sigma^2}{\sigma}\right), \quad \beta = \frac{1}{2}\frac{\partial^2 \omega_0}{\partial k_0^2}, \quad (A1)$$

$$\alpha = \omega_0 k_0^2 \frac{k_0^2 h^2 (9\sigma^8 - 28\sigma^6 + 38\sigma^4 - 28\sigma^2 + 9) + 2k_0 h\sigma(3\sigma^6 - 23\sigma^4 + 13\sigma^2 - 9) - 7\sigma^6 + 38\sigma^4 + 9\sigma^2}{16\sigma^4 \left(k_0^2 h^2 (1-\sigma^2)^2 - 2k_0 h\sigma(1+\sigma^2) + \sigma^2\right)}.$$

It is interesting to note that in the mentioned range of depths the wave frequency differs from the deep-water value in less than 0.1%; the linear phase velocity $\omega_0/k_0$ and the linear group velocity $C_{gr}$ differ from the corresponding deep-water limit values (3) in less than 1%. At the same time the effect of finite depth on the wave nonlinearity and dispersion is in fact much stronger. The dependences of nonlinear and dispersion coefficients, $\alpha$ and $\beta$, normalized by the values at the infinite depth, $\alpha_\infty$ and $\beta_\infty$, are shown in Fig. 14a. The conditions of experiments by Slunyaev et al. (2013) are marked with symbols.

It is constructive to reformulate the effect of finite water depth within the NLS framework (1) in terms of scaling of the wave amplitude and time. For arbitrary given water depth $h$ and solution of equation (1) $A(\xi, t)$ (where the commoving coordinate is introduced, $\xi = x - C_{gr}t$) the new scaled function

$$A'(\xi, \tau) = \mu A(\xi, t), \quad \text{where} \quad \mu = \sqrt{\frac{\alpha}{\beta}} \bigg/ \sqrt{\frac{\alpha_\infty}{\beta_\infty}}, \quad \tau = \beta t \qquad (A2)$$

satisfies the equation in form

$$i\frac{\partial A'}{\partial \tau} + \frac{\partial^2 A'}{\partial \xi^2} + \frac{\alpha_\infty}{\beta_\infty}|A'|^2 A' = 0. \qquad (A3)$$

In (A2), (A3) the coefficients with subscripts "∞" correspond to the infinitively deep water values given in (3). Equation (A3) does not depend on $h$ and thus is universal (meanwhile we assume $\alpha\beta > 0$, what is true for sufficiently deep water, $k_0h > 1.363$). Therefore if one considers a given initial condition $A(x, t = 0)$ in a basin of depth $h$, this problem is equivalent to consideration of the modified initial condition $\mu A(x, t = 0)$ in infinitively deep water (the time of evolution will be scaled though according to (A2)). Thus, the problem of a finite water depth is reduced to the initial value problem for conditions with modified amplitude, considered in the previous subsection. We may note that the water depth still enters the relation between $A$ and velocity potential $\phi$ (2) by virtue of $\omega_0$, but the potential is not used for determination of the wavemaker signal, and thus does not affect the conclusion.

The dependence of the scaling coefficient $\mu$ as function of the dimensionless depth as determined in (A2) is given in Fig. 14b, where the conditions of the laboratory experiments by Slunyaev et al. (2013) are shown by symbols. It is interesting to note that the case of the deepest water $k_0h = 5.7$ in fact corresponds to a 10% deficit of the wave amplitude, what is significant, since the soliton amplitude should drop down by 20%. The case $k_0h = 3.57$ corresponds to a more than 20% decrease of the wave amplitude. The initial condition is effectively twice smaller and will not produce a soliton if $k_0h < 2$.

**Table 1.** Conditions of laboratory experiments 1-4 with single or two similar solitary groups. The frequencies and dimensionless crest and trough amplitudes are calculated from the measured time series (lab) and from the numerical simulations (num).

| Exp. No | $\omega_p$, rad/s | | $\omega_m$, rad/s | | $h\omega_m^2/g$ | $\omega_m^2/gA_{cr}$ | | $\omega_m^2/gA_{tr}$ | | Setup | Description |
|---|---|---|---|---|---|---|---|---|---|---|---|
| | lab | num | lab | num | | lab | num | lab | num | | |
| 1 | 7.0 | 6.9 | 7.2 | 7.3 | 5.3 | 0.30 | 0.33 | 0.24 | 0.25 | 1 | strong transverse modulation |
| 2 | 5.8 | 5.9 | 6.0 | 6.1 | 3.7 | 0.23 | 0.30 | 0.19 | 0.23 | 1 | reflection |
| 3 | 5.6 | 5.7 | 6.0 | 6.1 | 3.7 | 0.32 | 0.39 | 0.25 | 0.28 | 1 | reflection |
| 4 | 5.9 | 5.9 | 6.1 | 6.1 | 3.7 | 0.29 | 0.30 | 0.23 | 0.23 | 2 | reflection and head-on collision |

**Table 2.** Conditions of laboratory experiments 5-7 with two solitary groups. The parameters are calculated for data of the numerical simulations.

| Exp. No | $\omega_m$, rad/s | | $k_m$, rad/m | | $\omega_m^2/gA_{cr}$ | | $\omega_m^2/gA_{tr}$ | | $V$, m/s | | Setup | Description |
|---|---|---|---|---|---|---|---|---|---|---|---|---|
| | 1 | 2 | 1 | 2 | 1 | 2 | 1 | 2 | 1 | 2 | | |
| 5 | 6.1 | 6.1 | 3.5 | 3.3 | 0.30 | 0.39 | 0.23 | 0.28 | 3.03 | 3.16 | 3 | reflection and head-on collision |
| 6,7 | 7.1 | 5.9 | 4.9 | 3.4 | 0.19 | 0.18 | 0.16 | 0.16 | 3.62 | 2.97 | 4 | over-head collision |

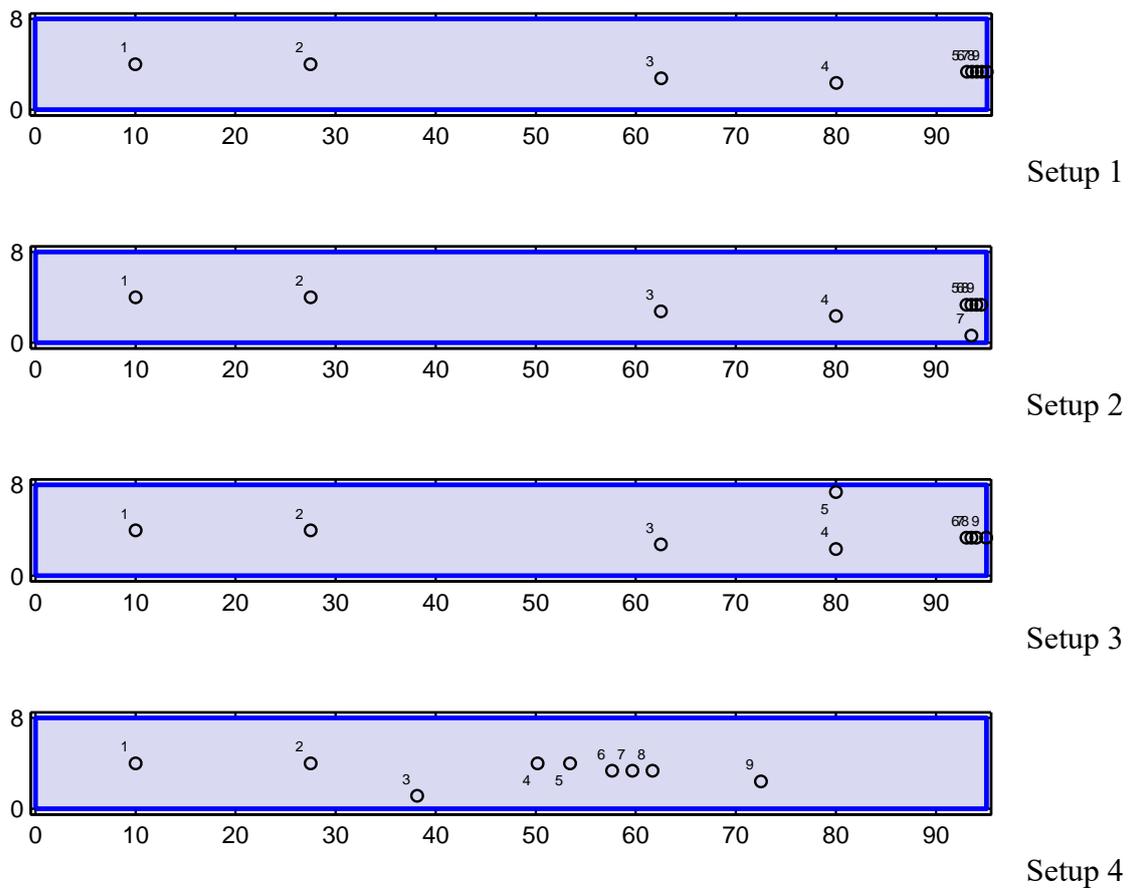

Fig. 1. Locations of probes in the laboratory experiments. The wavemaker is situated at the left side.

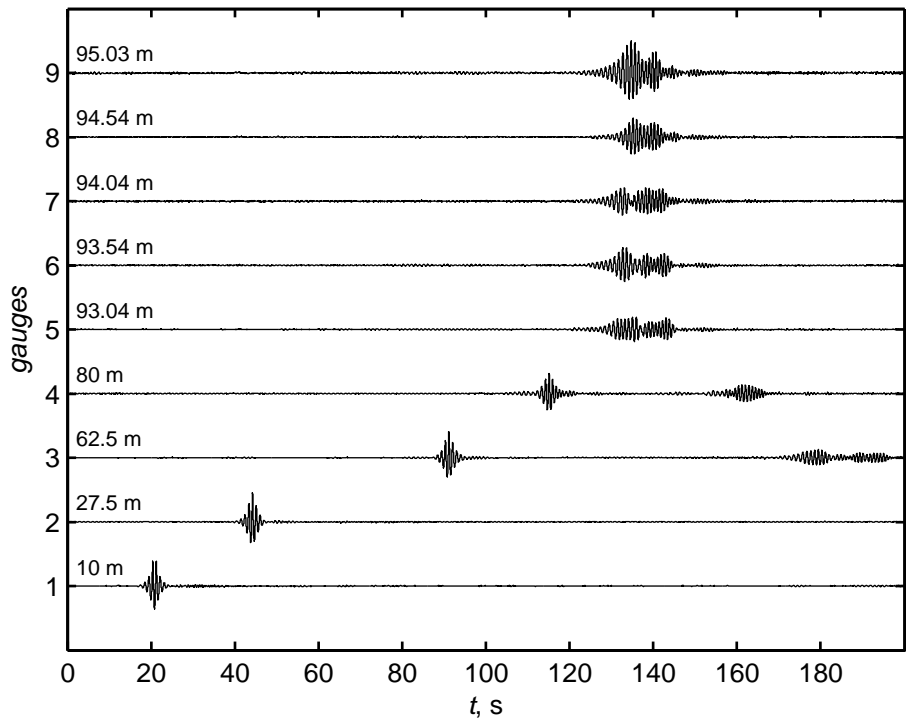

Exp. 1

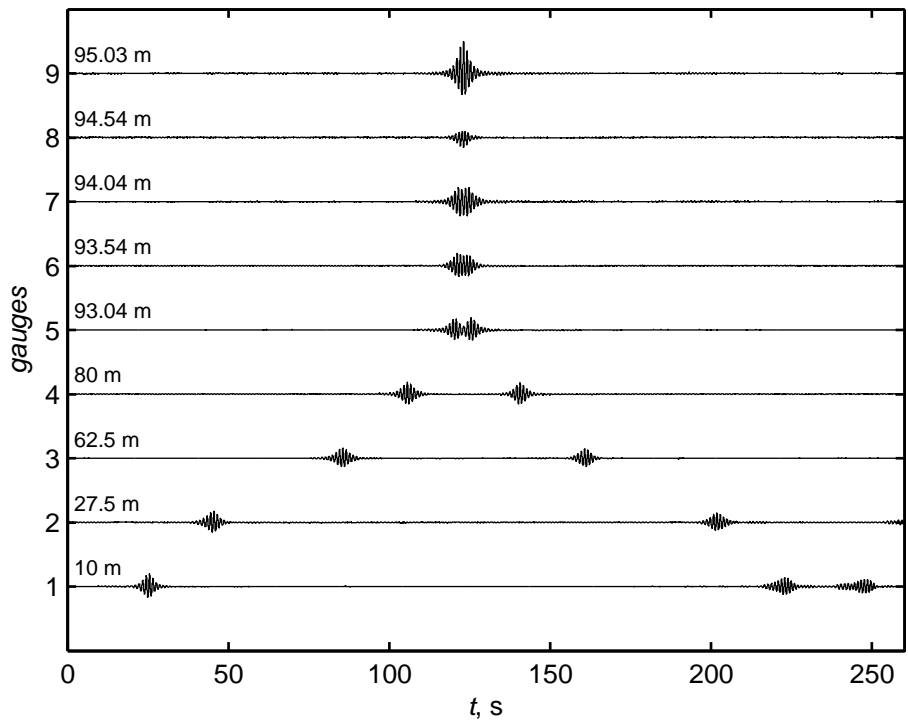

Exp. 2

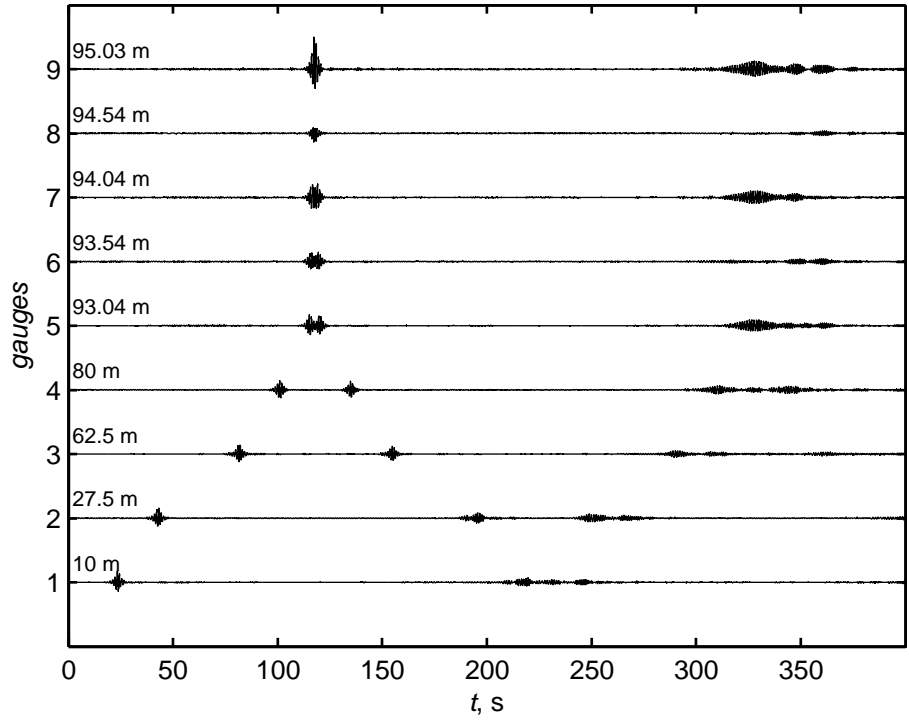

Exp. 3

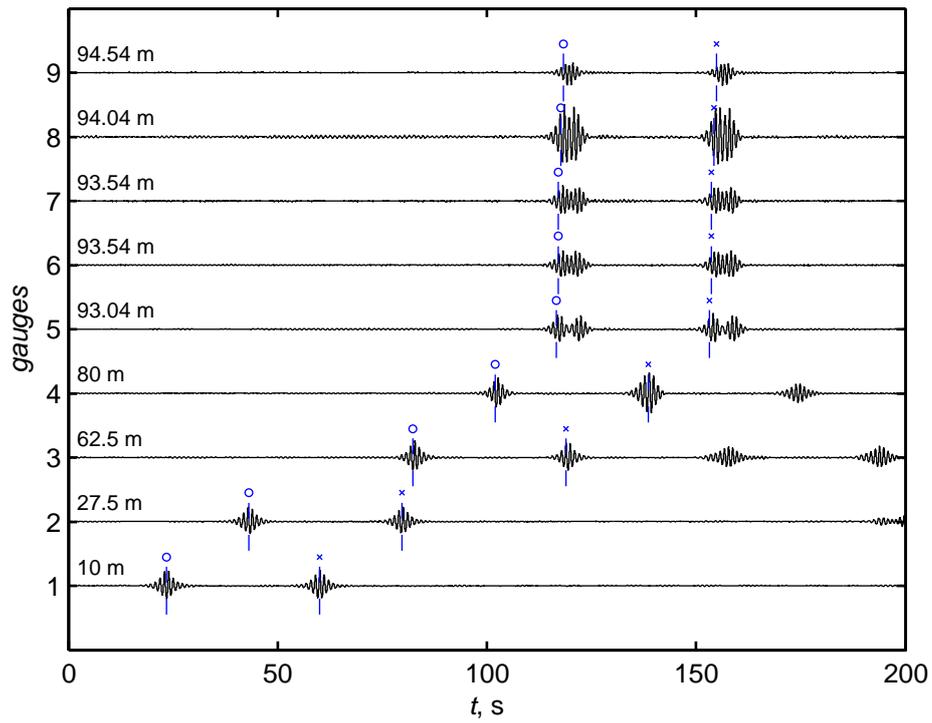

Exp. 4

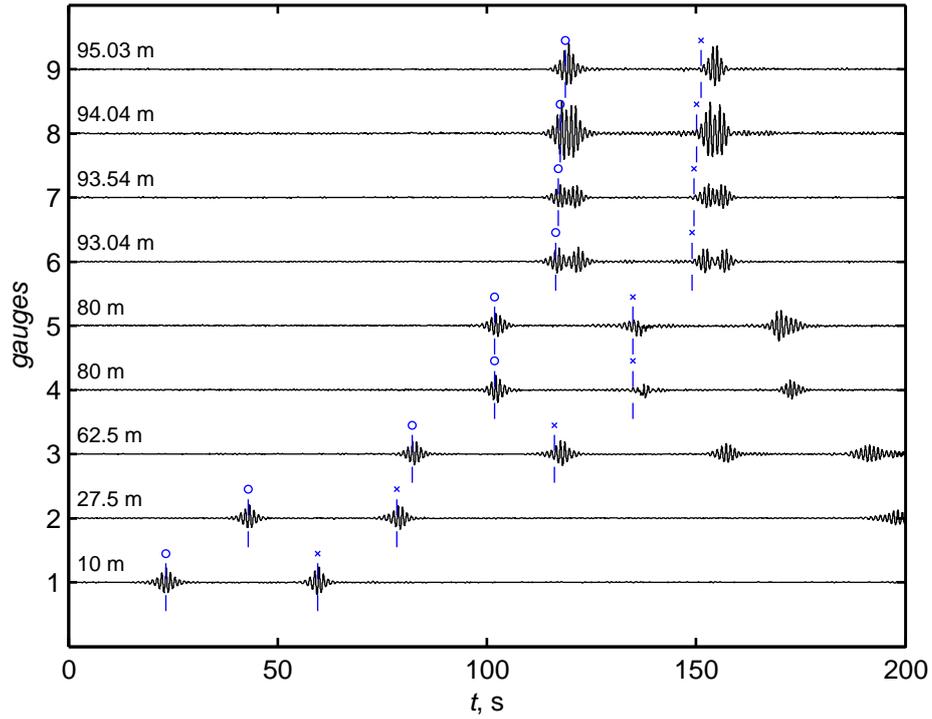

Exp. 5

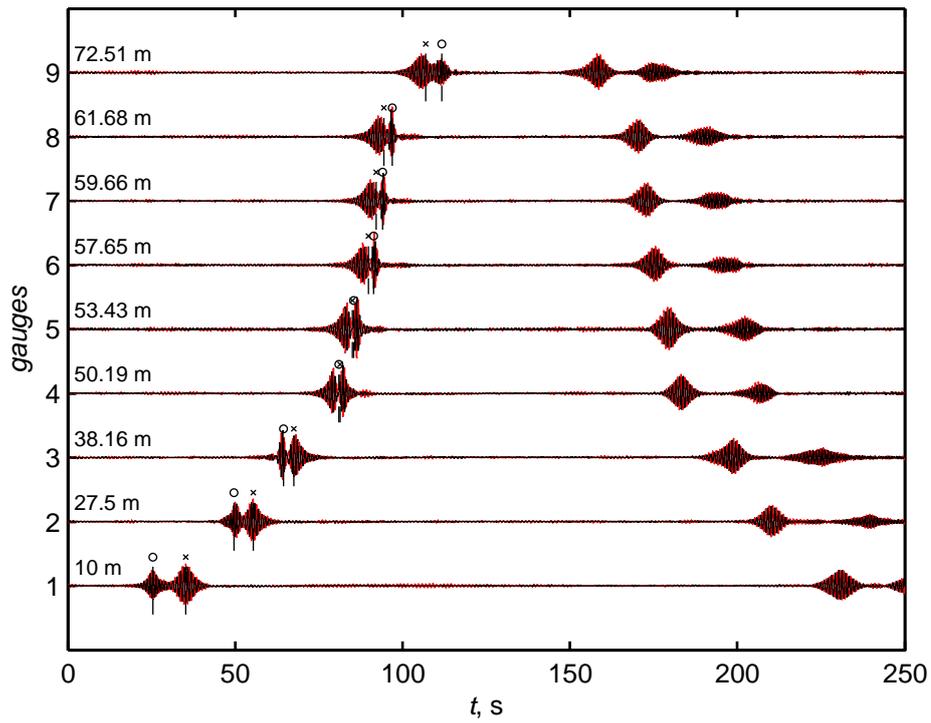

Exp. 6&7

Fig. 2. Time series of the surface displacements in the laboratory experiments. The distance from the wavemaker is indicated by numbers. See also Fig. 1 for locations of the probes.

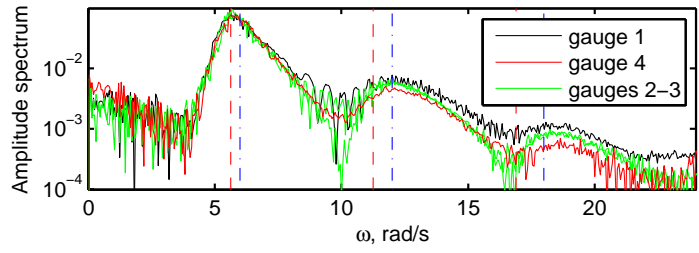
Exp. 3

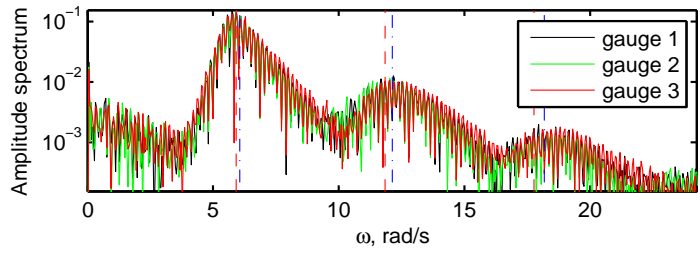
Exp. 4

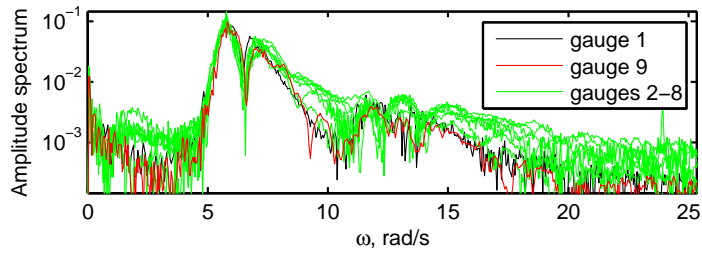
Exp. 6

Fig. 3. Amplitude frequency spectra of the solitary wave groups recorded in the laboratory experiments.

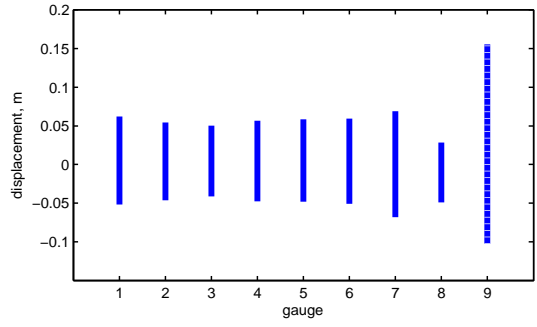
Exp. 2

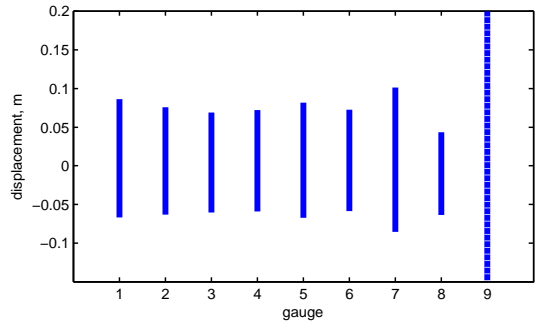
Exp. 3

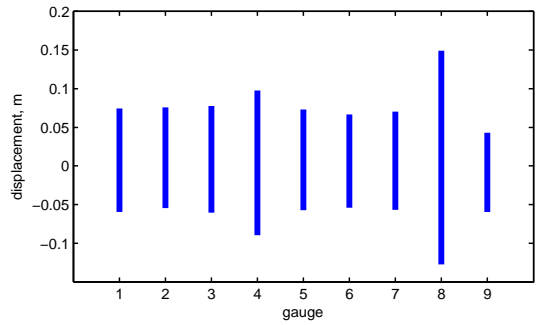
Exp. 4

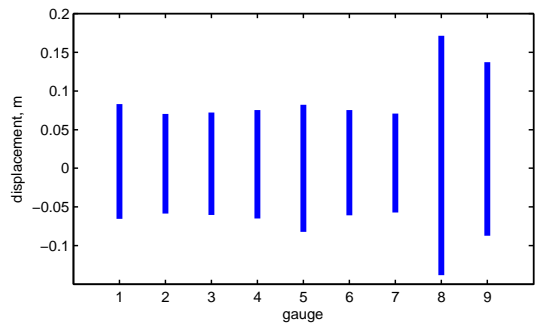
Exp. 5

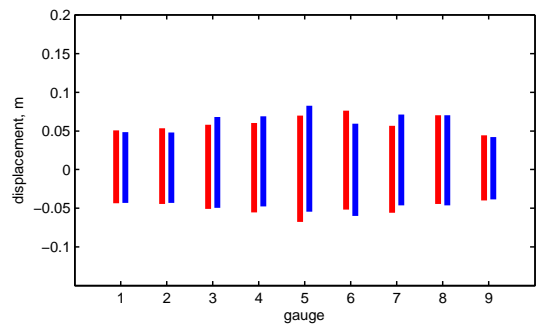
Exp. 6&7

Fig. 4. Ranges of the recorded surface displacement at different probes. The data from the gauge 9 in Exp. 2, 3 may be with bias (shown with interrupted bands).

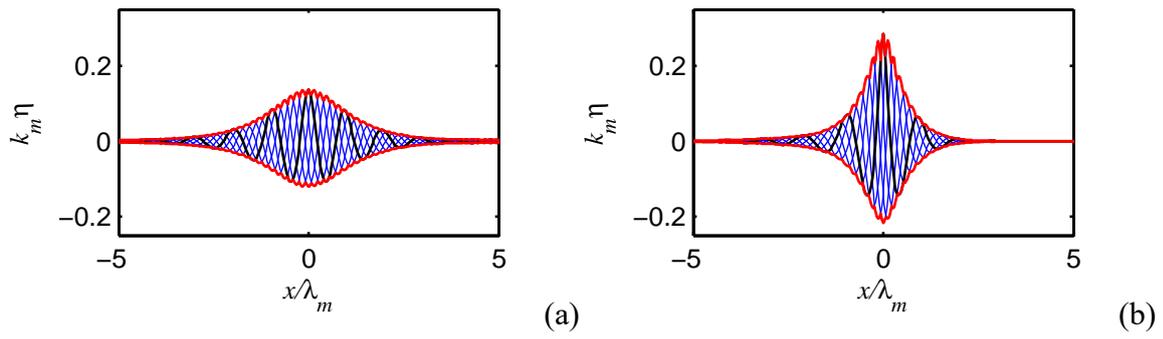

Fig. 5. Solitary wave groups used in the numerical tests: with the steepness $k_m A_{cr} \approx 0.14$ (a) and $k_m A_{cr} \approx 0.30$ (b). The red curves show the envelope, while blue and black curves correspond to the displacements at six different phases.

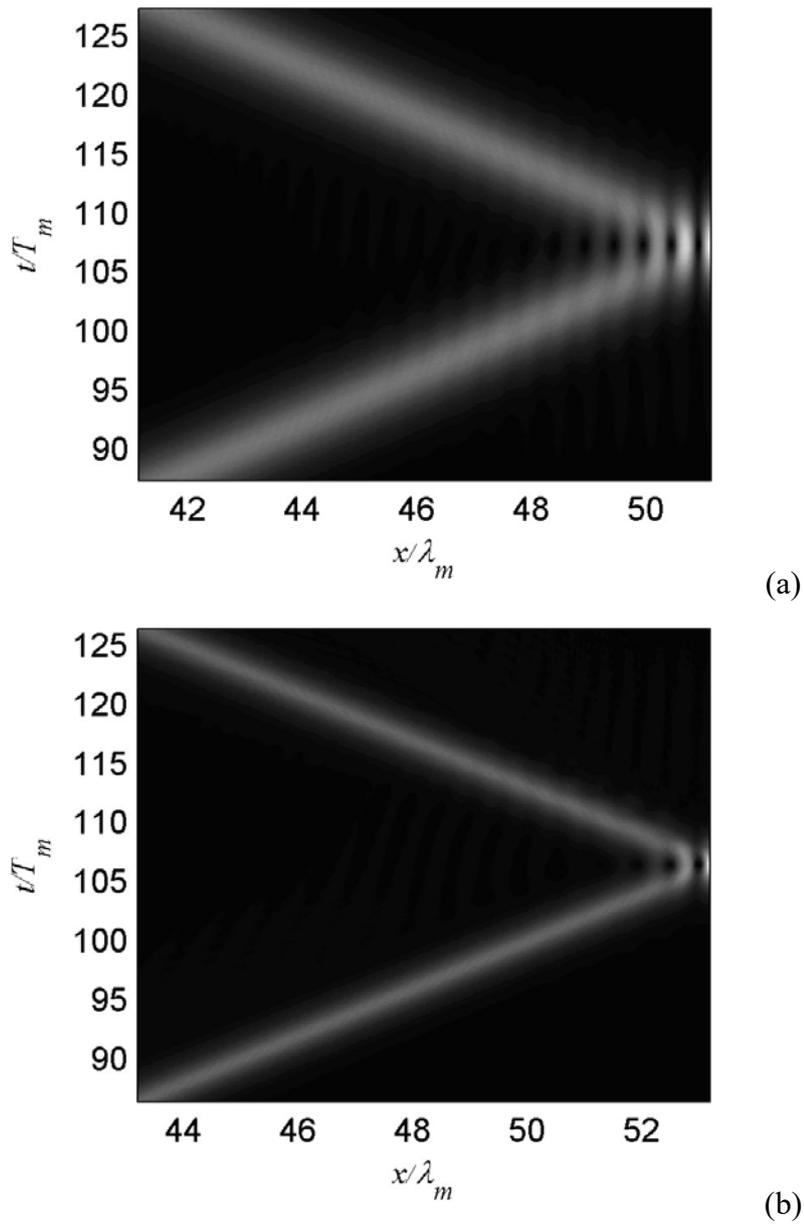

Fig. 6. Space-time diagrams of the upper wave envelope in the numerical simulations of the reflection of solitons with moderate ($k_m A_{cr} \approx 0.14$) (a) and large ($k_m A_{cr} \approx 0.30$) (b) steepness. Only the area near the wall is shown.

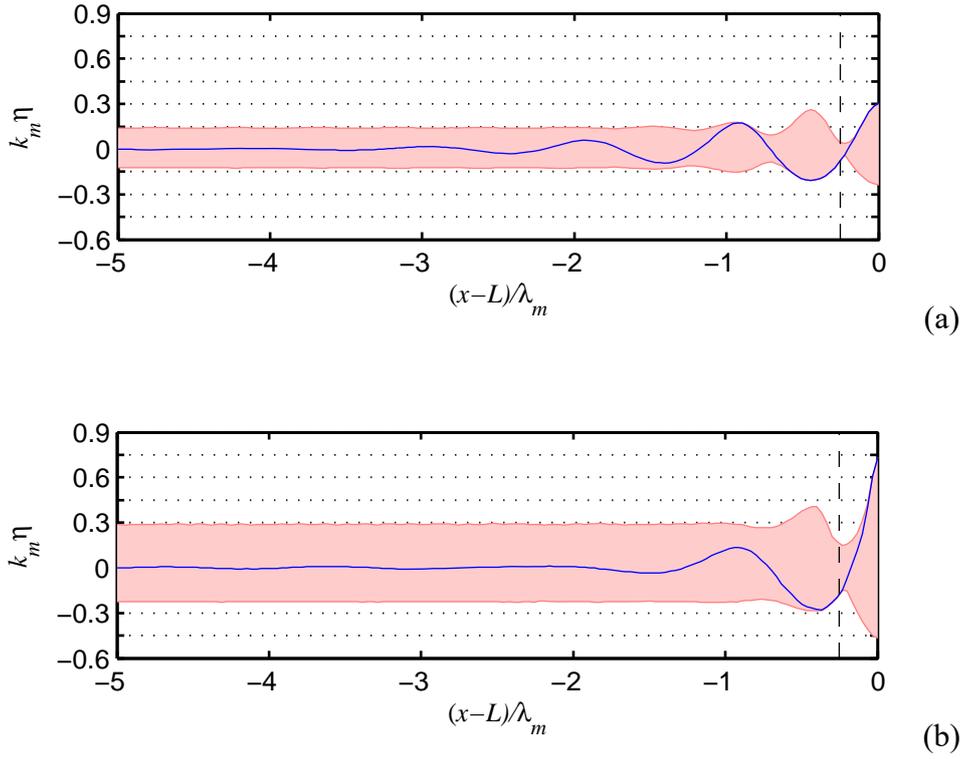

Fig. 7. Maximum waves in the numerical simulations of reflecting solitary groups with the steepness $k_m A_{cr} \approx 0.14$ (a) and $k_m A_{cr} \approx 0.30$ (b). The limits of the surface displacement in the simulated domain are shown with shading as functions of space. The vertical dashed lines correspond to distance $\lambda_m/4$ from the wall.

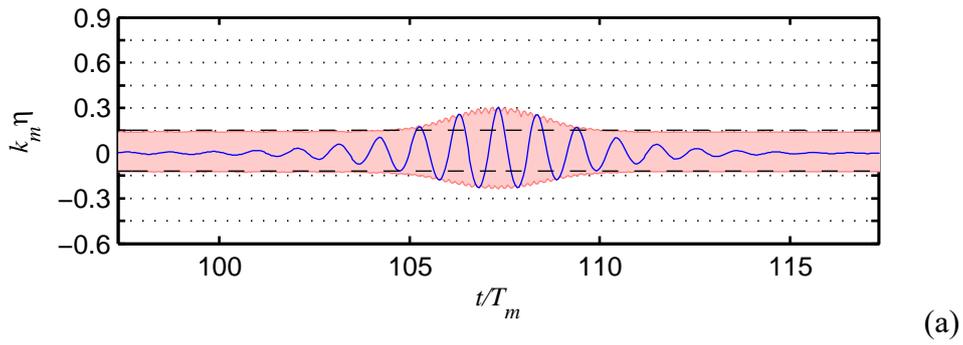

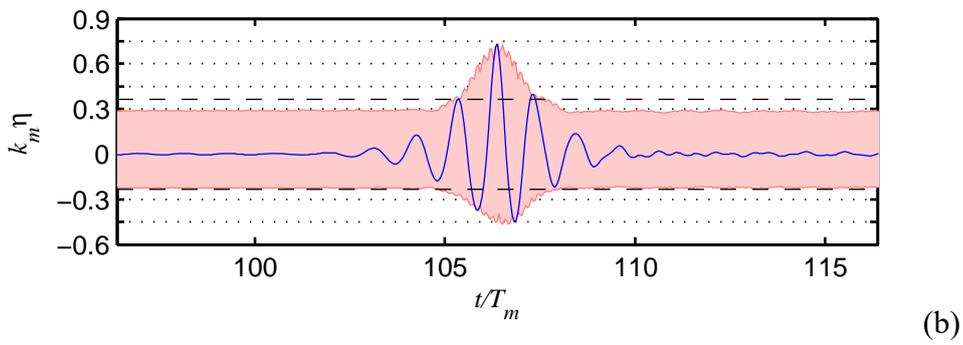

Fig. 8. Same as in Fig. 7, but the time series. The horizontal broken lines show the levels of the 1/2 of the maximum water rise/depression attained in the simulation.

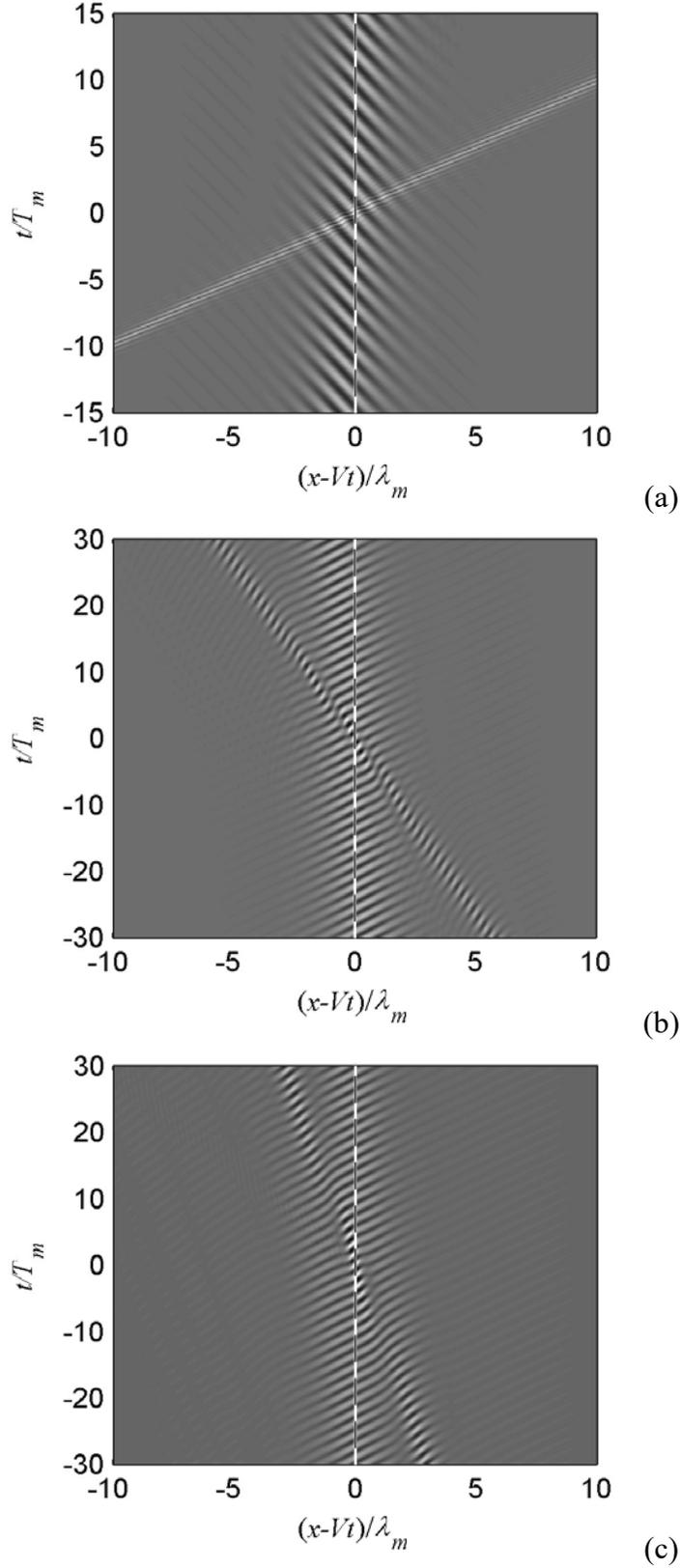

Fig. 9. Surface displacement as a function of time and space in the numerical simulations of the head-on (a) and over-head (b) collisions of envelope solitons with parameters $k_m A_{cr} \approx 0.30$, $\omega_m \approx 8.5$ rad/s and $k_m A_{cr} \approx 0.14$, $\omega_m \approx 6$ rad/s; and of the over-head collision of solitons $k_m A_{cr} \approx 0.30$, $\omega_m \approx 7$ rad/s and $k_m A_{cr} \approx 0.14$, $\omega_m \approx 6$ rad/s (c). The reference wave period, frequency and velocity are of the second soliton.

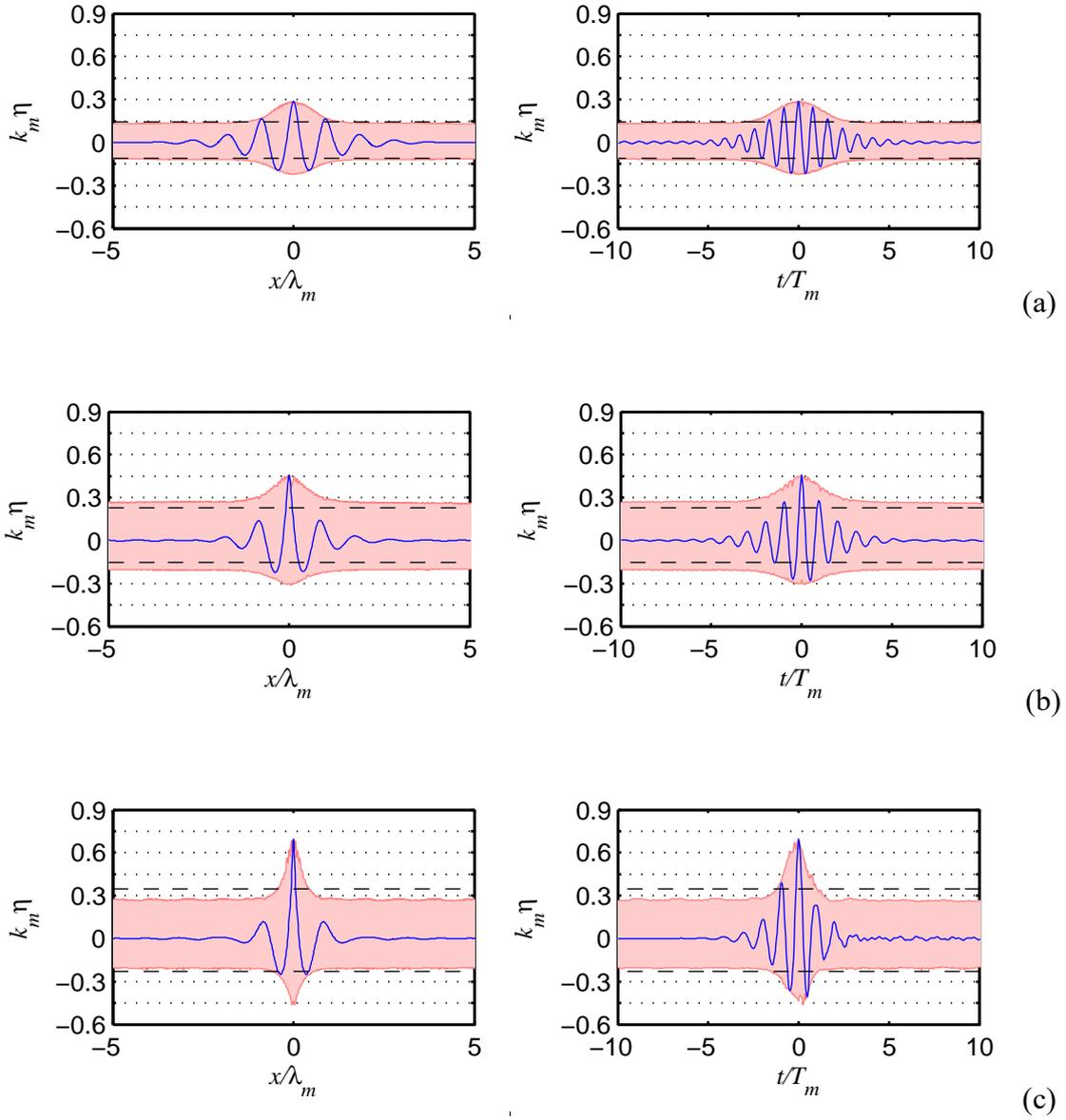

Fig. 10. Head-on collisions of solitary groups with the same carrier frequency $\omega_m \approx 6$ rad/s, the snapshots (left column) and time series (right column): two similar solitons with $k_m A_{cr} \approx 0.14$ (a), solitons $k_m A_{cr} \approx 0.30$ and $k_m A_{cr} \approx 0.14$ (b), and two solitons with steepness $k_m A_{cr} \approx 0.30$ (c). The surface displacements which correspond to the maximum waves are given by thin lines. The limits of the surface displacement in the simulated domain as functions of space/time are shown with shading. The horizontal broken lines show levels of the 1/2 of the maximum water rise/depression attained in the simulation.

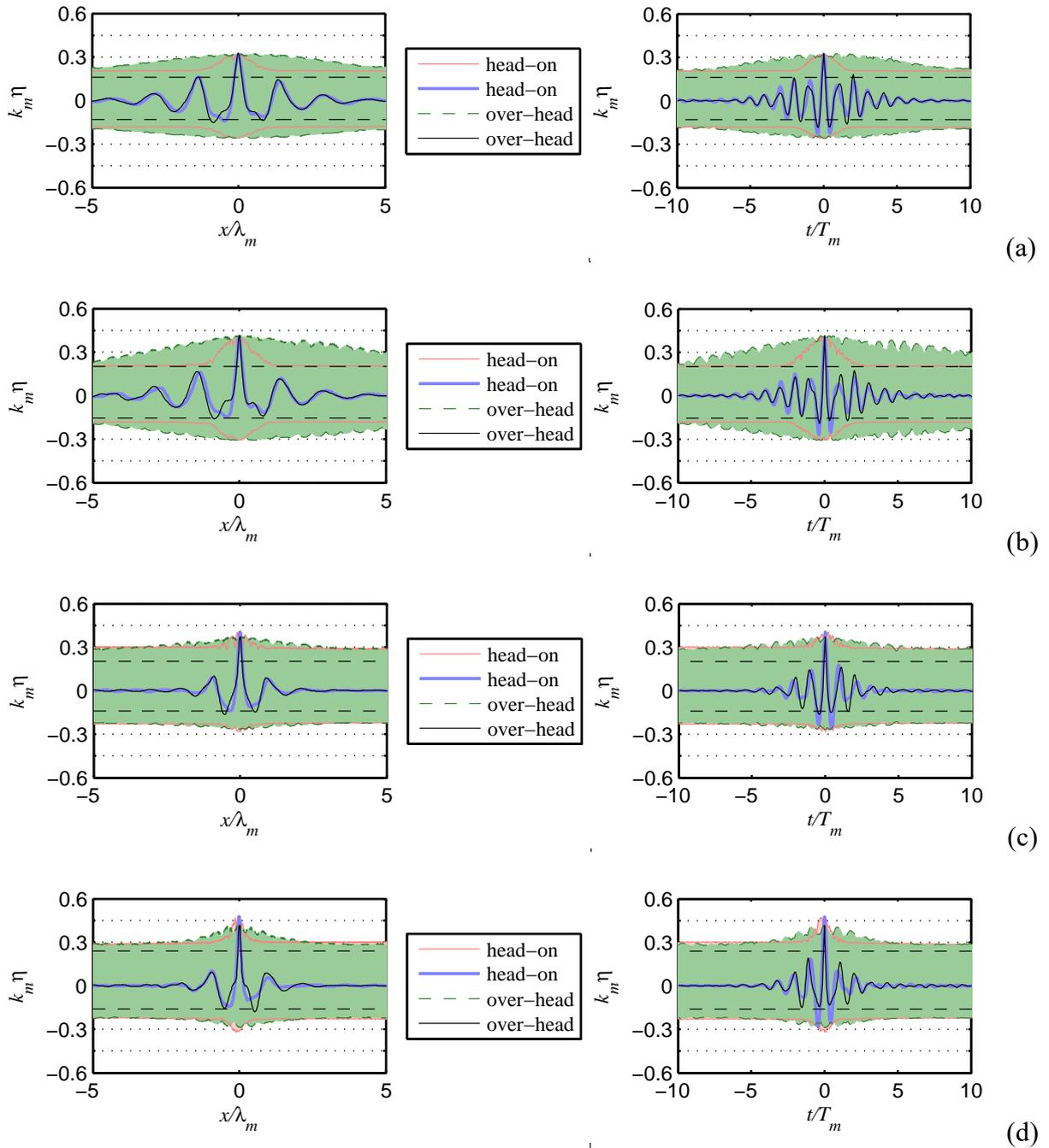

Fig. 11. Collisions of solitary groups with different frequencies of the carrier, the snapshots (left column) and time series (right column): solitons with the same steepness $k_m A_{cr} \approx 0.14$ (a), solitons $k_m A_{cr} \approx 0.30$, $\omega_m \approx 8.5$ rad/s and $k_m A_{cr} \approx 0.14$, $\omega_m \approx 6$ rad/s (b), solitons with inverse combination of steepnesses $k_m A_{cr} \approx 0.14$, $\omega_m \approx 8.5$ rad/s and $k_m A_{cr} \approx 0.30$, $\omega_m \approx 6$ rad/s (c) and steep solitons $k_m A_{cr} \approx 0.30$ (d). The surface displacements which correspond to the maximum waves are given by thick blue (head-on collisions) and thin black (over-head collisions) curves. The limits of the surface displacement in the simulated domain as functions of space/time are also shown by shadings, see the legend. The horizontal broken lines show levels of the 1/2 of the maximum water rise/depression attained in the simulation. The reference wavenumber, wave period and frequency correspond to the long wave.

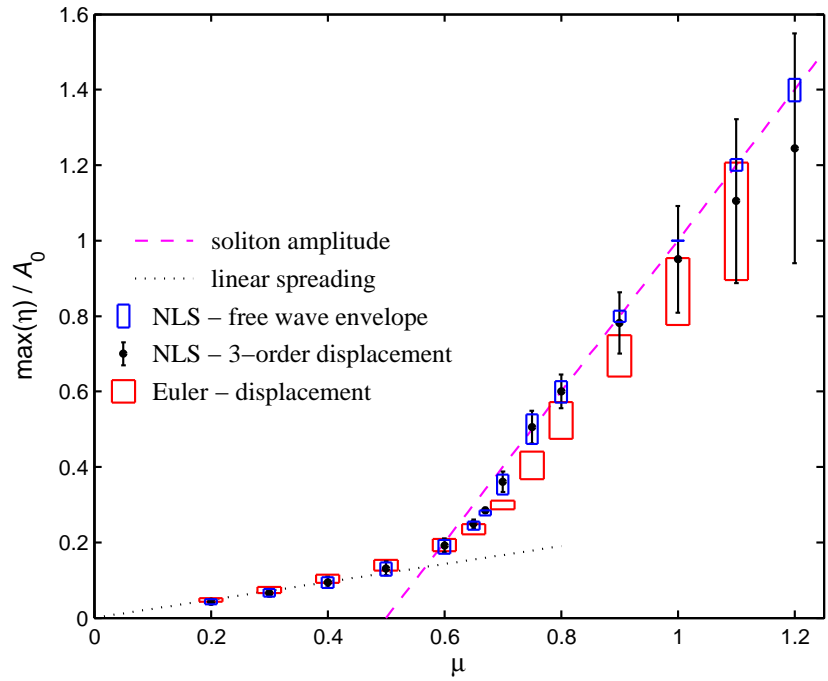

Fig. 12. Maximum displacements in the long-term solution of the Cauchy problem for the initial condition in form of the NLS envelope soliton with modified amplitude: the frameworks of the Euler equations and the NLS equation (see the legend). The analytic estimations are given by lines. The unperturbed soliton steepness is $a = 0.25$, the amplitude factor, $\mu$, varies from 0.2 to 1.2.

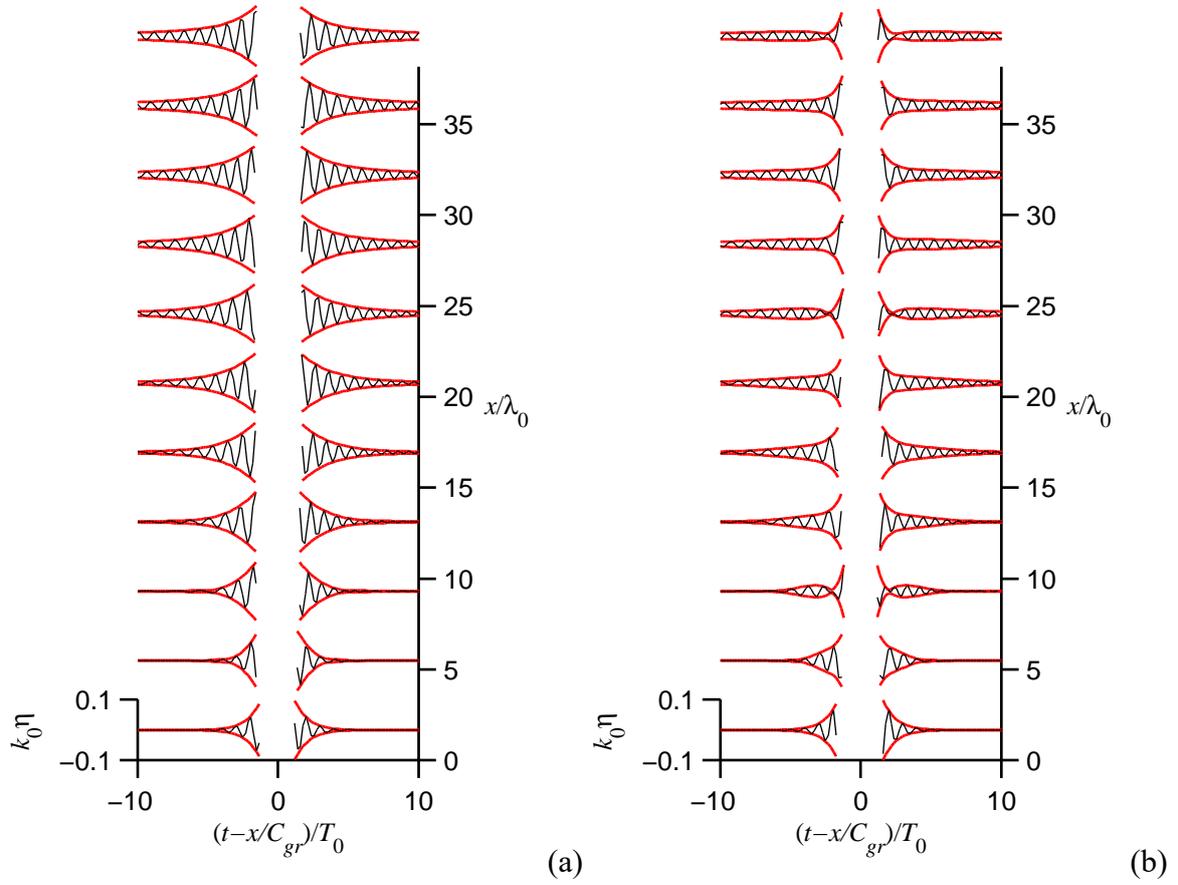

Fig. 13. Simulations of the NLS envelope soliton with steepness $a = 0.2$ with amplitude deficit, $\mu = 0.8$, (a), and excess, $\mu = 1.2$, (b). The spatial form of the NLS equation is simulated, the time series of the surface displacement and the envelopes $|A|$ are shown in co-moving references; cropping at $\pm 0.1$ is applied to show the tails.

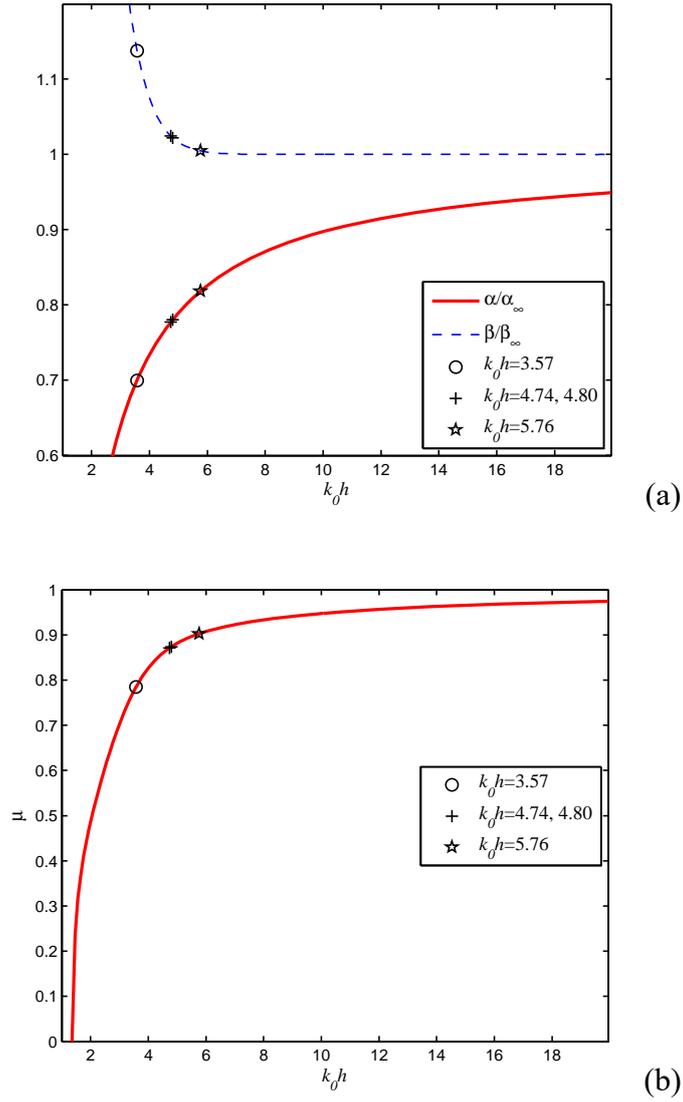

Fig. 14. Coefficients of nonlinearity and dispersion of the NLS equation (a), and the effective scaling coefficient $\mu$ (b) as functions of the dimensionless water depth $k_0 h$. The conditions of laboratory experiments by Slunyaev et al. (2013a) are shown with symbols.